\documentclass[preprint]{aastex}

\newcommand\kms           {km~s$^{-1}$}

\newcommand{\tnm}         {\tablenotemark}
\newcommand{\tnt}         {\tablenotetext}
\newcommand{\hii}         {\ion{H}{2}}
\newcommand{\sk}          {\\[0.5ex]}
\newcommand{\ns}          {\\*[-0.5ex]}
\newcommand{\nolin}       {0.\phn\phn&\nodata}
\newcommand{\nss}         {\\}
\newcommand{\source}[1]   {\\[-1.0ex] \tableline \\*[-2.0ex] \multicolumn{10}{c}{#1} \\
*[-2.0ex] \\* \tableline}
\newcommand{\fontfortable}{\fontsize{7}{10}\selectfont}
\shorttitle{Full-Polarization OH in Massive SFRs: Data}
\shortauthors{Fish et al.}

\begin{document}

\title{Full-Polarization Observations of OH Masers in Massive
  Star-Forming Regions: I. Data}
\author{Vincent L.~Fish\altaffilmark{1}}
\affil{National Radio Astronomy Observatory}
\affil{P. O. Box O, 1003 Lopezville Road, Socorro, NM  87801}
\email{vfish@nrao.edu}
\author{Mark J.~Reid and Alice L.~Argon}
\affil{Harvard--Smithsonian Center for Astrophysics}
\affil{60 Garden Street, Cambridge, MA  02138}
\email{reid@cfa.harvard.edu, aargon@cfa.harvard.edu}
\and
\author{Xing-Wu Zheng}
\affil{Department of Astronomy, Nanjing University}
\affil{Nanjing 210093  PR China}
\email{xwzheng@nju.edu.cn}

\altaffiltext{1}{Jansky Fellow}

\begin{abstract}
We present full-polarization VLBA maps of the ground-state, main-line,
$^2\Pi_{3/2}, J = 3/2$ OH masers in 18 Galactic massive star-forming
regions.  This is the first large polarization survey of interstellar
hydroxyl masers at VLBI resolution.  A total of 184 Zeeman pairs are
identified, and the corresponding magnetic field strengths are
indicated.  We also present spectra of the NH$_3$ emission or
absorption in these star-forming regions.  Analysis of these data will
be presented in a companion paper.
\end{abstract}


\section{Introduction}
\doublespace
Hydroxyl (OH) masers are an important tool for probing the environment
of high-mass star-forming regions (SFRs).  They are bright, often
greater than 1~Jy, so they can be seen in SFRs that are located across
the Galaxy.  Due to Zeeman splitting, the magnetic field strength and
line-of-sight direction can be inferred \emph{in situ}.  The linear
polarization direction and fraction provides additional information
about the three-dimensional alignment of the magnetic field.  The
velocity field of many points around the central star can also be
probed using the velocity of emission of maser features.

Only one high-mass SFR has been studied in great depth: W3(OH).
Images obtained using VLBI techniques conclusively detected the
presence of Zeeman pairs of OH maser spots in all four ground-state
($^2\Pi_{3/2}, J = 3/2$) transitions \citep{lo75,reidw3,fouquet82} as
well as in the 6035 MHz $^2\Pi_{3/2}, J = 5/2$ transition
\citep{moran78}.  Subsequent imaging uncovered a plethora of Zeeman
pairs in the ground-state transitions \citep{garcia,w3oh} as well as
in the two main-line $^2\Pi_{3/2}, J = 5/2$ transitions
\citep{desmurs98} and the 13441 MHz $^2\Pi_{3/2}, J = 7/2$
transition \citep{baudry98}; additionally, W3(OH) is known to have a
Zeeman pair at 13434 MHz \citep{gusten94}.
More recently, \citet{wright04a,wright04b} have mapped all four
ground-state transitions in full polarization.

These observations led to several surprising results.  First, maser
spots appeared to be clustered on a scale of several times
$10^{15}$~cm \citep{reidw3,wright04a}.  Second, the magnetic fields
implied by Zeeman splitting consistently had the same line-of-sight
direction across W3(OH), suggesting an organized magnetic field
structure.  Third, proper motions obtained over an eight-year time
baseline showed that the OH maser spots were moving outward from the
central ultracompact (UC) \hii\ region \citep{w3oh}.
\citet{wright04a} conclude from a third epoch of data taken a decade
later that the OH motions trace rotation in the north-south direction
as well as expansion, a conclusion supported by the distribution of
4765 MHz $^2\Pi_{1/2}, J = 1/2$ masers as well
\citep{harveysmith05}.  Fourth, most maser spots were 100\% polarized,
and most of these were nearly totally circularly polarized
\citep{garcia}.  No spots with a linear polarization fraction greater
than 45\% were detected.  Fifth, maser distributions in one transition
correlate with those in certain transitions but not with others.  For
instance, the spatial distributions of 1720 and 4765 MHz
masers overlap \citep{baudry88,gray01,palmer03}, while the 1665 and
1667 MHz transitions are spatially separated \citep{norris81}.

While OH masers have been observed using VLBI techniques in other
massive SFRs \citep[for instance,][]{zrm,arm2,slysh}, the total number
of such sources observed remains only a handful.  In order to more
rigorously test the universality of the properties discovered in
W3(OH), as well as in the hope of uncovering additional interesting
phenomena, we have undertaken a survey of 18 sources, including two
for which preliminary results were presented by \citet{arm2}.

\section{Observations and Data Reduction}

\singlespace
\begin{deluxetable}{lcc@{ }c@{ }cc@{ }c@{ }ccccc}
\tabletypesize{\footnotesize}
\tablecaption{Observed Sources\label{obs_tab}}
\tablehead{
  \colhead{} & \colhead{} & \multicolumn{3}{c}{} &
  \multicolumn{3}{c}{} & \colhead{Dist.} &
  \colhead{}  & \colhead{Obs.} &
  \colhead{Distance}\\
  \colhead{Source} & \colhead{Alias} & \multicolumn{3}{c}{RA} &
  \multicolumn{3}{c}{Dec} & \colhead{(kpc)} &
  \colhead{Epoch\tnm{a}}  & \colhead{Time\tnm{b}} &
  \colhead{Reference}\\
}
\startdata
G005.886$-$0.393 & \nodata & 18&00&30.41 & $-$24&04&00.9 &  3.8     & 4 & 135 & 1 \\
G009.622$+$0.195 & \nodata & 18&06&14.74 & $-$20&31&33.9 &  5.7     & 4 & 135 & 2 \\
G010.624$-$0.385 & \nodata & 18&10&28.68 & $-$19&55&49.7 &  4.8     & 4 & 135 & 1 \\
G034.257$+$0.154 & \nodata & 18&53&18.67 & $+$01&14&58.5 &  3.8     & 3 & 450 & 3 \\
G035.577$-$0.029 & \nodata & 18&56&22.50 & $+$02&20&27.1 & 10.5\phn & 4 & 150 & 1 \\
G040.622$-$0.137 & \nodata & 19&06&01.61 & $+$06&46&35.8 &  2.2     & 4 & 150 & 4 \\
G043.796$-$0.127 & \nodata & 19&11&54.01 & $+$09&35&50.0 &  9.0     & 4 & 135 & 5 \\
G049.488$-$0.387 & W51 M/S & 19&23&43.93 & $+$14&30&31.5 &  7.0     & 1 & 500 & 6 \\
G069.540$-$0.976 & ON 1    & 20&10&09.05 & $+$31&31&35.2 &  3.0     & 2 & 220 & 7 \\
G070.293$+$1.601 & K3$-$50 & 20&01&45.73 & $+$33&32&45.3 &  8.7     & 2 & 220 & 8 \\
G075.782$+$0.343 & ON 2 N  & 20&21&43.97 & $+$37&26&38.1 &  5.6     & 2 & 220 & 9 \\
G081.721$+$0.571 & W75 S   & 20&39&00.96 & $+$42&22&48.1 &  2.0     & 2 & 220 &10 \\
G081.871$+$0.781 & W75 N   & 20&38&36.39 & $+$42&37&34.3 &  2.0     & 2 & 220 &10 \\
G109.871$+$2.114 & Cep A   & 22&56&17.87 & $+$62&01&48.6 &  0.7     & 3 & 240 &11 \\
G111.543$+$0.777 & NGC 7538& 23&13&45.34 & $+$61&28&10.1 &  2.8     & 3 & 240 &12 \\
G196.454$-$1.677 & S269    & 06&14&37.07 & $+$13&49&36.3 &  3.8     & 5 & 180 &13 \\
G213.706$-$12.60 & Mon R2  & 06&07&47.84 & $-$06&22&56.7 &  0.9     & 5 & 180 & 9 \\
G351.775$-$0.538 & \nodata & 17&26&42.70 & $-$36&09&17.4 &  2.2     & 1 & 255 &14
\enddata
\tablecomments{Units of right ascension are hours, minutes, and seconds,
and units of declination are degrees, arcminutes, and arcseconds.  All
coordinates are J2000.}
\tnt{a}{Epochs of observation: (1) 1996 March 01-02,
(2) 2000 November 22 and 2001 January 06,
(3) 2001 February 16-17,
(4) 2001 May 26 and 2001 May 28, (5) 2002 February 06.}
\tnt{b}{Approximate amount of total on-source observing time, in
minutes.}
\tablerefs{(1) \citealt{fish}; (2) \citealt{hofner};
(3) \citealt{reifenstein};
(4) \citealt{hughes}; (5) \citealt{watson}; (6) \citealt{genzel};
(7) \citealt{araya}; (8)
\citealt{harris75}; (9) \citealt{wink}; (10) \citealt{dickel}; 
(11) \citealt{blaauw59}; (12) \citealt{crampton78};
(13) \citealt{moffat}; (14) \citealt{caswell97}
}
\end{deluxetable}
\doublespace

We observed 18 sources in 5 separate epochs with the National Radio
Astronomy Observatory's\footnote{The National Radio Astronomy
Observatory is a facility of the National Science Foundation operated
under cooperative agreement by Associated Universities, Inc.} Very
Long Baseline Array (VLBA).  Sixteen of the sources were observed at
the frequencies of both main-line $^2\Pi_{3/2}, J = 3/2$ OH
transitions (1665.4018 and 1667.3590~MHz) using 125~kHz bands divided
into 128 spectral channels, which corresponds to a velocity width of
0.176 \kms\ per channel.  W51 and G351.775$-$0.538 were observed using
250~kHz bands instead, resulting in a velocity width of 0.352 \kms\
per channel.  Both the parallel and cross-polarization data were
correlated in order to produce images in all Stokes parameters.  Dates
of observations and on-source observing times can be found in Table
\ref{obs_tab}.

Data processing was done using the NRAO Astronomical Image Processing
System (AIPS).  A full description of the calibration procedure occurs
in Appendix A.  The source 3C286 was used to calibrate polarization
position angles.  For each source, an iterative self-calibration was
performed.  A single maser spot in a single maser transition and
polarization was selected based on overall intensity on all baselines,
and the resulting self-calibration was applied to both transitions and
polarizations of the source.  In sources for which interstellar
scatter broadening is large, extended baselines were discarded.  Since
interstellar scattering enlarges the observed angular size of objects,
the visibilities on long baselines (corresponding to small angular
scales) are nearly zero.  These baselines are therefore not useful,
since their visibilities simply add noise to the images.  Due to the
distribution of VLBA antennas, this is effectively equivalent to
discarding all data from the Mauna Kea and St.~Croix antennas for
moderate scatter broadening (e.g., in G40.622$-$0.137); for large
scatter broadening (e.g., in K3$-$50), the Hancock, North Liberty, and
Brewster data were discarded as well.  The resulting synthesized beam
size for each source is listed in Tables \ref{g5t} to \ref{g351t}.

Imaging was performed using the AIPS task \texttt{IMAGR}.  First,
low-resolution Stokes I map cubes were made in both transitions
encompassing an area greater than that of known OH maser emission
\citep{arm}.  Then, all fields in this map that contained OH
maser emission in one or more velocity planes were imaged with a
cellsize approximately one-fifth the synthesized beam size.  Maser
spots were identified in each velocity plane of these cubes.
Gaussians were fit to the maser spots using a modified version of the
AIPS task \texttt{JMFIT}.

Maser lines were identified when maser emission was seen in
essentially the same location in two or more consecutive velocity
planes.  Zeeman pairing of maser lines in LCP and RCP was done in a
similar fashion to the method used in \citet{fram}, but the
milliarcsecond resolution afforded by the VLBA permits unambiguous
identification of components of a Zeeman pair in the vast majority of
cases.  When an LCP line and an RCP line at a different velocity were
found in the same region of space typically to within $10^{15}$~cm,
they were identified as a Zeeman pair.  This separation was chosen
based on the empirical results of previous observations of W3(OH)
\citep{reidw3}.

Additionally, we observed the $(J,K) = (1,1)$ line of ammonia (rest
frequency 23694.495~MHz) with the VLA.  Observations of
G5.886$-$0.393, G9.622$+$0.195, and G351.775$-$0.538 were taken on
2001 September 27 with the VLA in DnC configuration, while the rest of
our sources were observed on 2002 January 6 in the D configuration.
On-source observation times were about 35 minutes for the first set of
sources and 20 minutes for the second set.  Data reduction was carried
out using AIPS.  Gaussians were fitted to the spectra where absorption
or emission was detected.

\section{Images and Tables\label{images}}

\begin{figure}
\begin{center}
\includegraphics[width=6.0in]{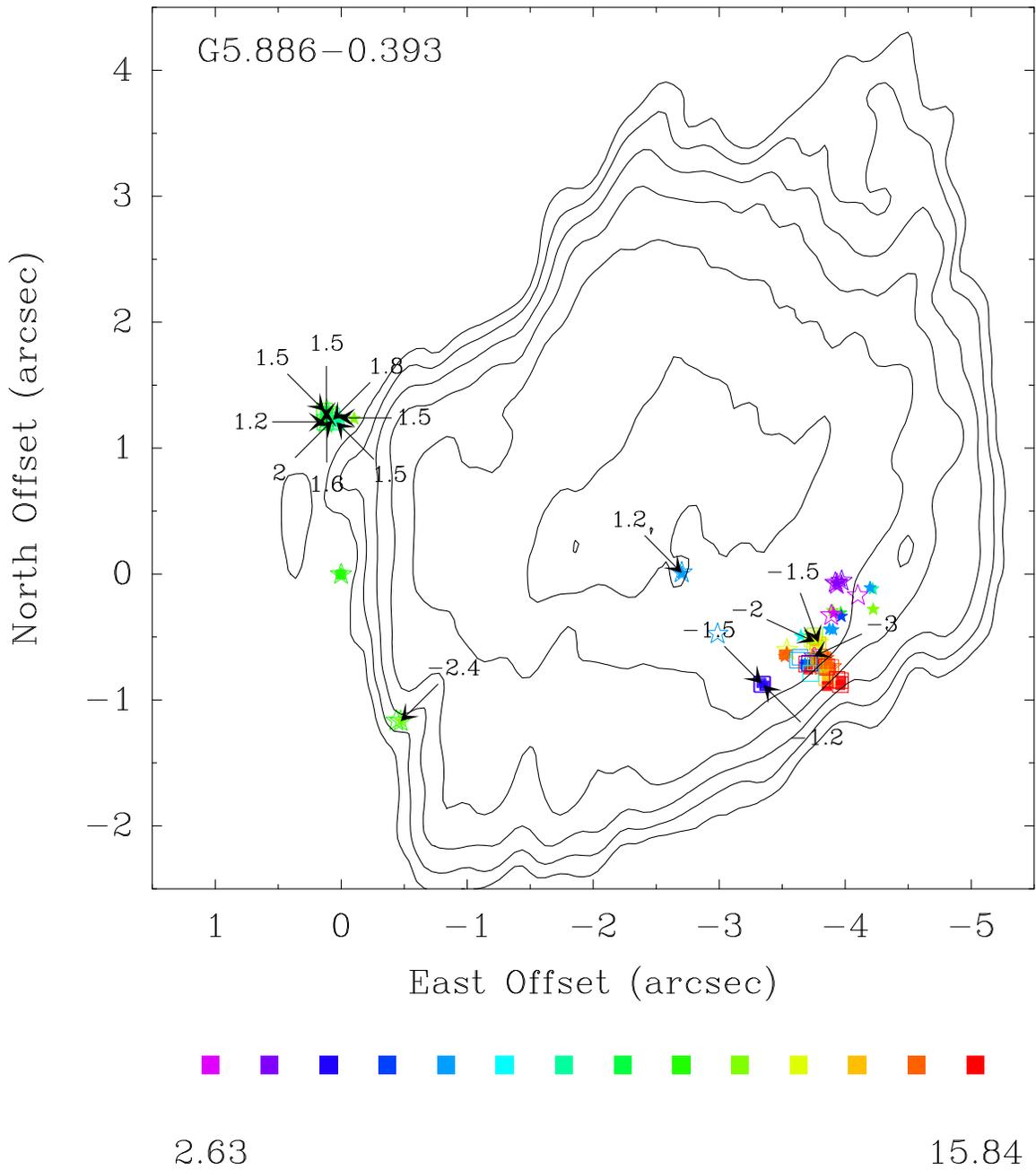}
\end{center}
\singlespace
\caption[Plot of maser spots in G5.886$-$0.393.]{Plot of maser spots
in G5.886$-$0.393.  Stars (squares) represent 1665 (1667) MHz OH
masers; open (filled) represents RCP (LCP) spots.  Numbers indicate
magnetic fields in milligauss, with a positive sign indicating that
the line-of-sight field direction is oriented away from the Sun.  The
masers are superposed on an X-band continuum map.  Contours start at
four times the rms noise and increase by factors of two.  The color
scale indicates LSR maser velocities in \kms.\label{g5v}}
\doublespace
\end{figure}

\begin{figure}
\begin{center}
\includegraphics[width=6.0in]{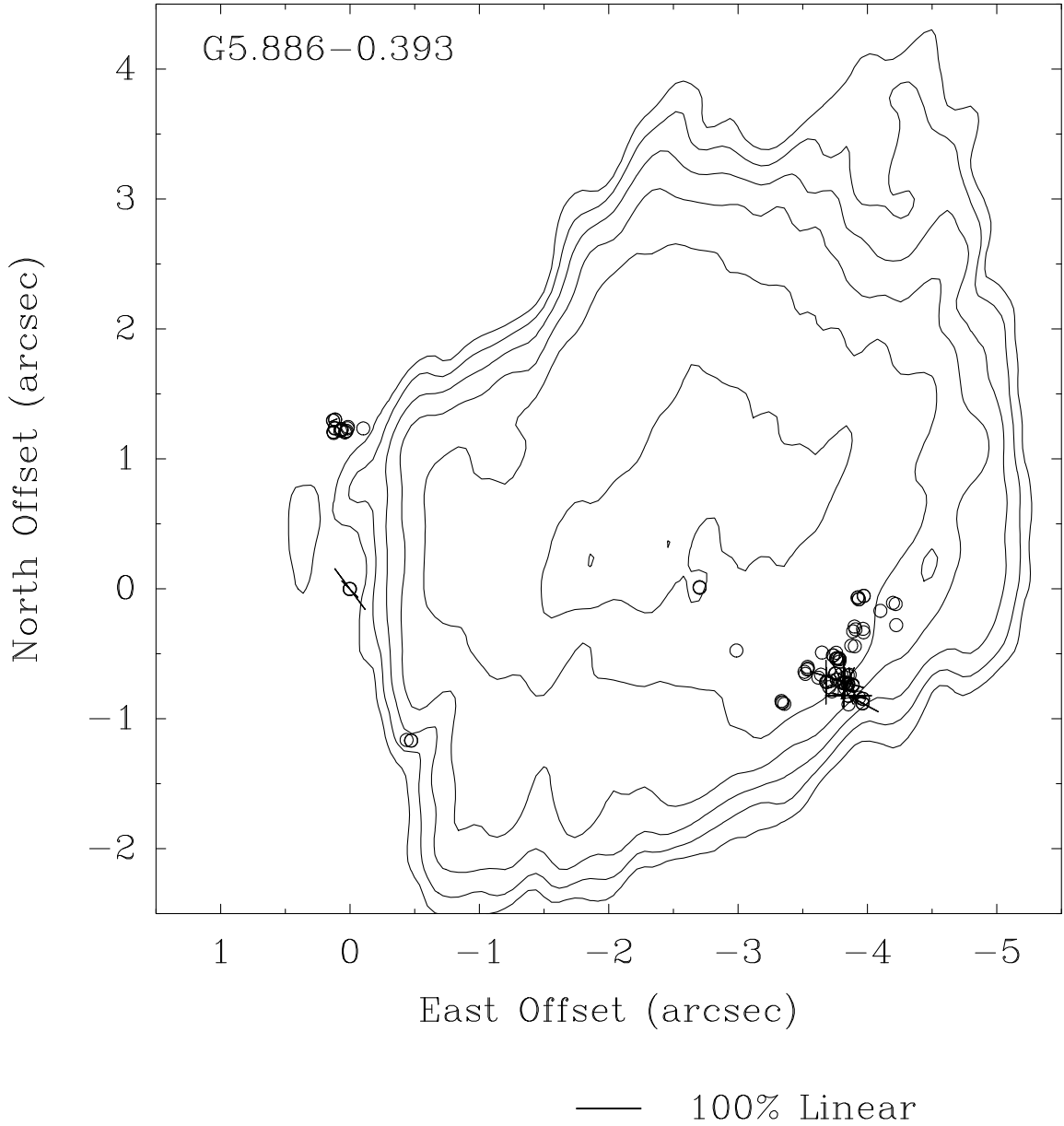}
\end{center}
\singlespace
\caption[Polarization map of G5.886$-$0.393.]{Polarization map of
G5.886$-$0.393.  Maser spots are represented as circles.  Lines
running through the centers of the circles indicate the linear
polarization fraction and direction.  A bar below the box indicates
the length of the polarization vector that would be drawn for a spot
that is totally linearly polarized.  The X-band continuum map is shown
for reference.\label{g5p}}
\doublespace
\end{figure}

\begin{figure}
\begin{center}
\includegraphics[width=6.0in]{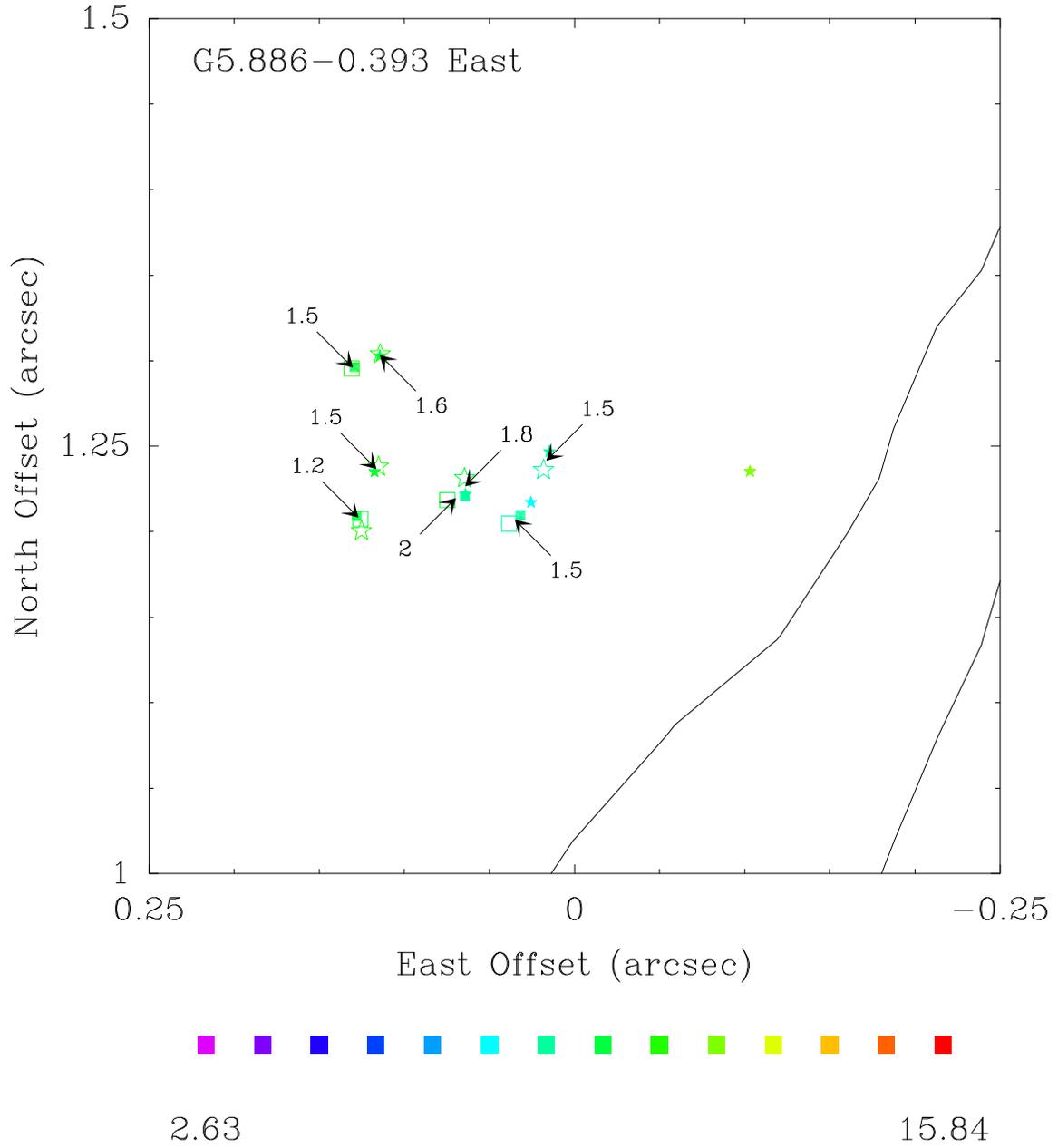}
\end{center}
\singlespace
\caption[Plot of maser spots in eastern cluster of
G5.886$-$0.393.]{Enlargement of plot of maser spots in the eastern
cluster of G5.886$-$0.393.  Symbols are as in Figure
\ref{g5v}.\label{g5ve}}
\doublespace
\end{figure}

\begin{figure}
\begin{center}
\includegraphics[width=6.0in]{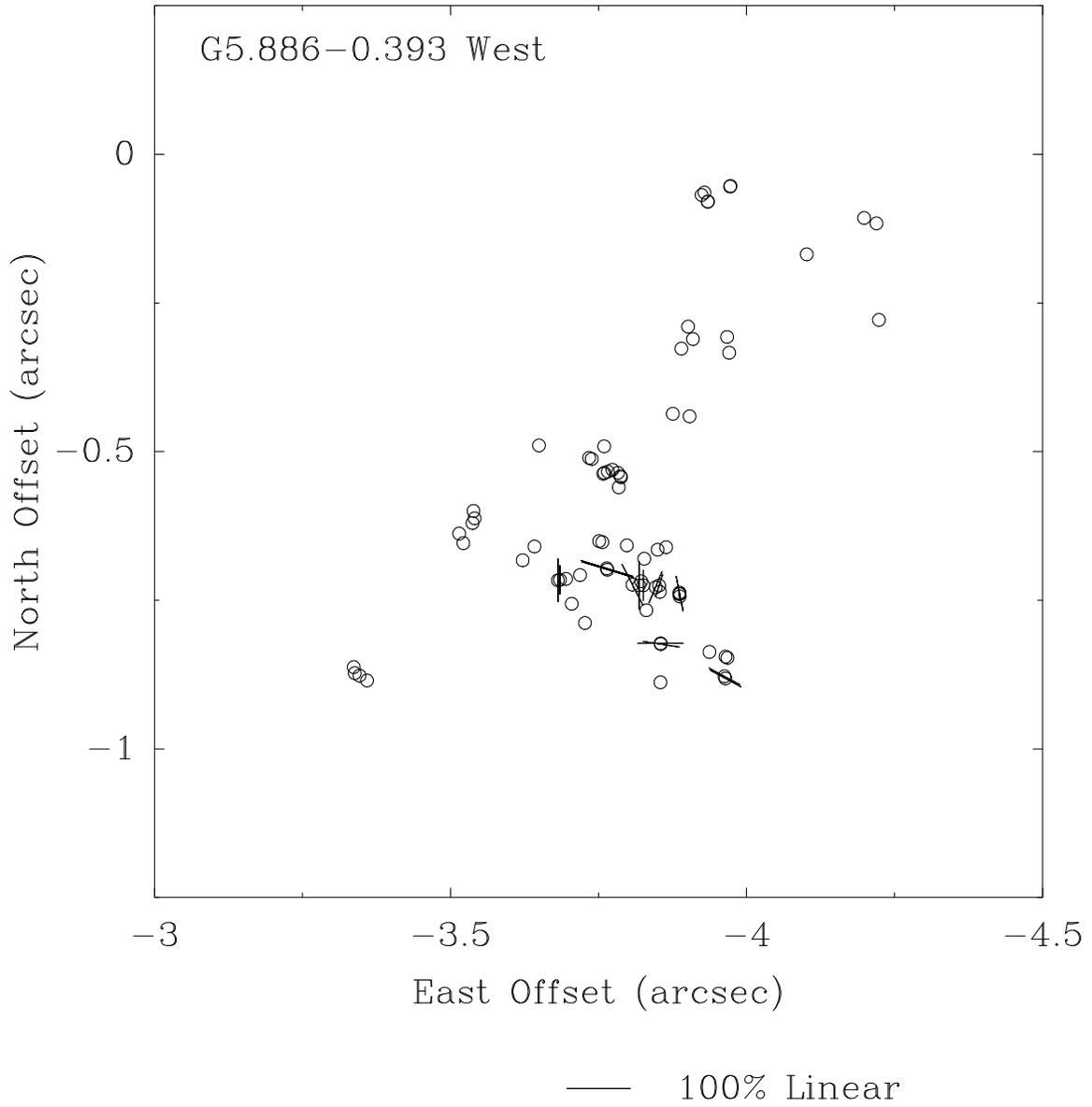}
\end{center}
\singlespace
\caption[Polarization map of western cluster in
G5.886$-$0.393.]{Enlargement of polarization map of the western
cluster of G5.886$-$0.393.  The continuum image is suppressed for
clarity.  Symbols are as in Figure \ref{g5p}.\label{g5pw}}
\doublespace
\end{figure}

\begin{figure}
\begin{center}
\includegraphics[width=6.0in]{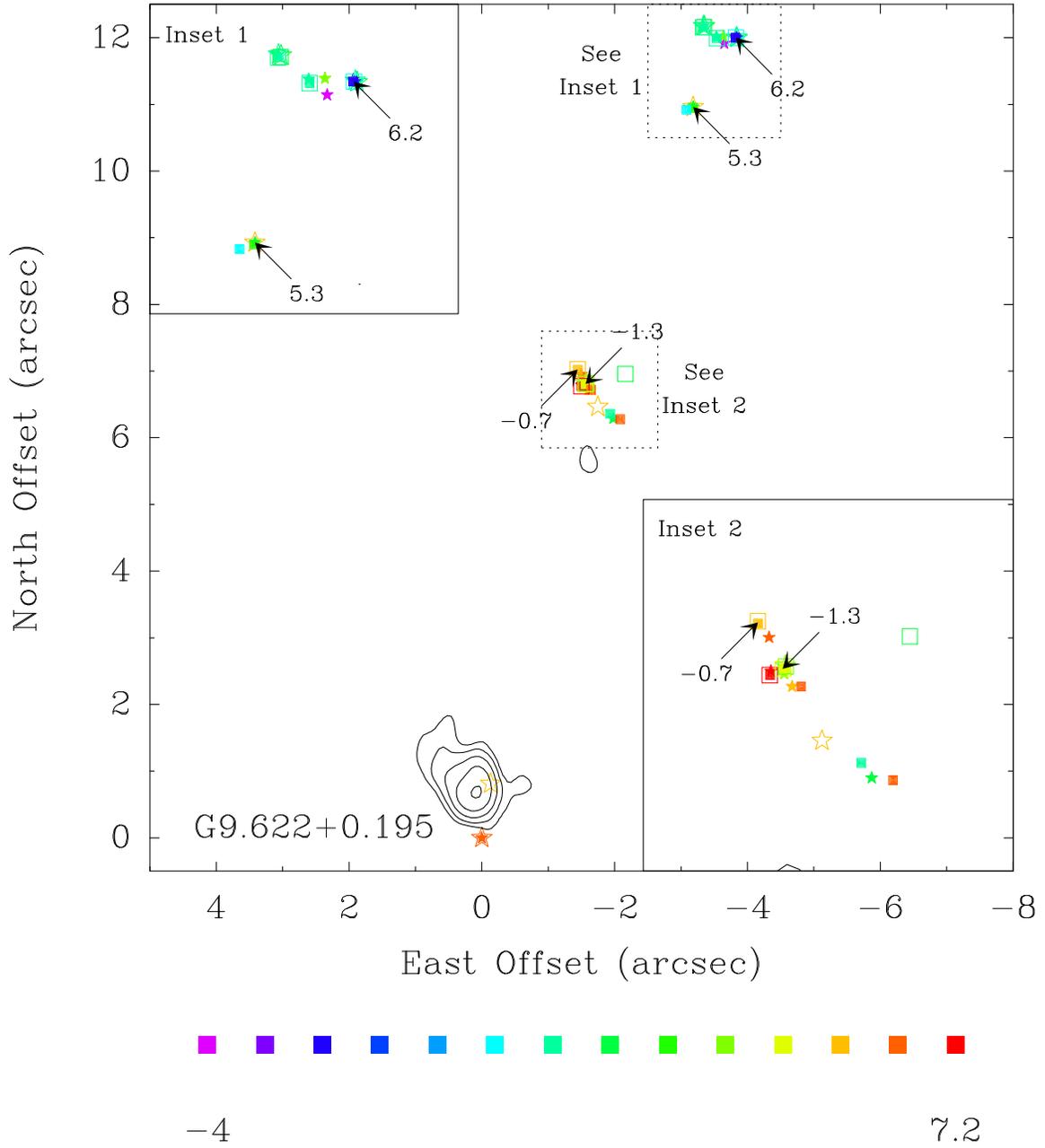}
\end{center}
\singlespace
\caption[Plot of maser spots in G9.622$+$0.195.]{Plot of maser spots
in G9.622$+$0.195.  Symbols are as in Figure \ref{g5v}.\label{g9v}}
\doublespace
\end{figure}

\begin{figure}
\begin{center}
\includegraphics[width=6.0in]{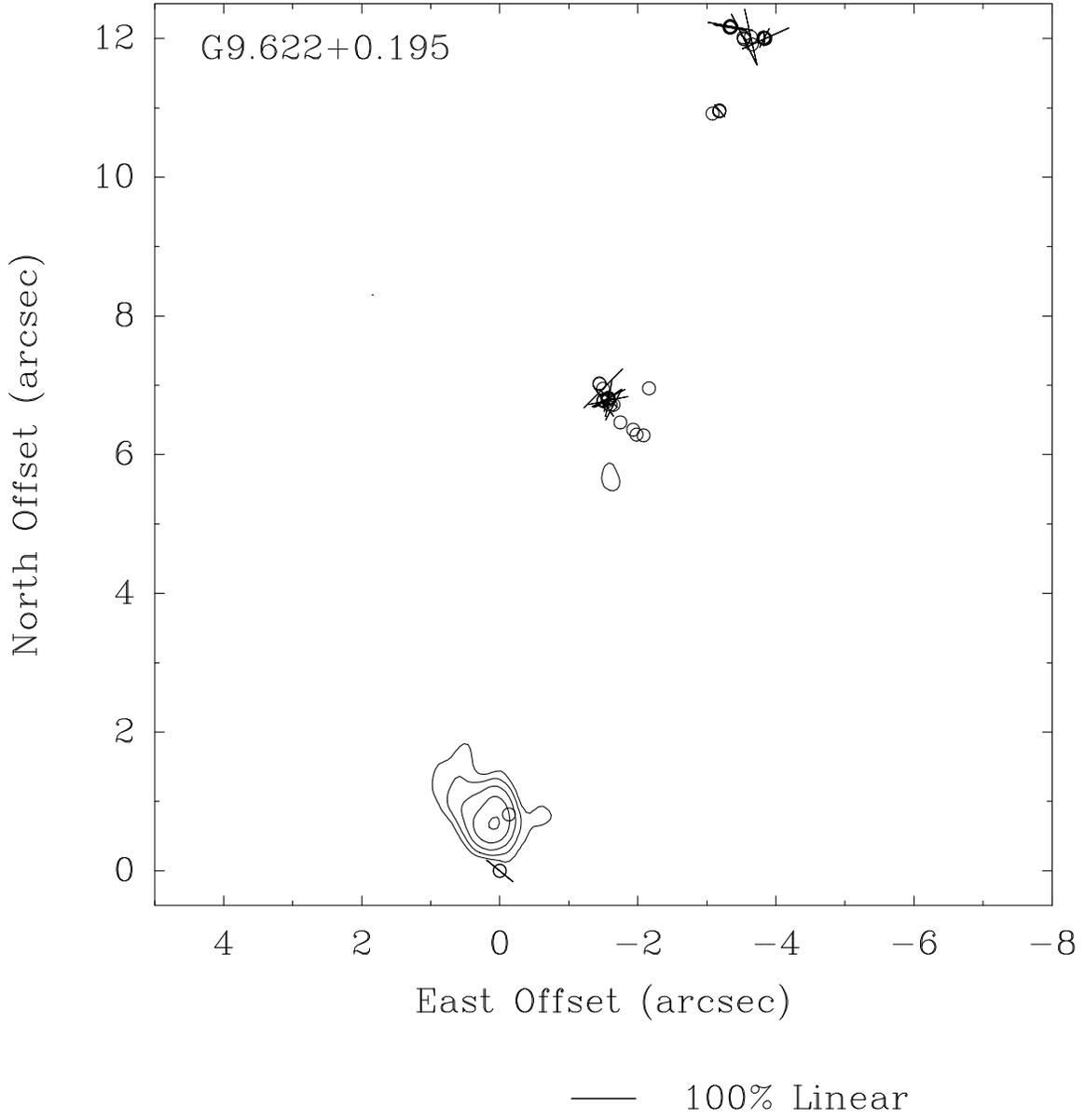}
\end{center}
\singlespace
\caption[Polarization map of G9.622$+$0.195.]{Polarization map of
G9.622$+$0.195.  Symbols are as in Figure \ref{g5p}.\label{g9p}}
\doublespace
\end{figure}

\begin{figure}
\begin{center}
\includegraphics[width=6.0in]{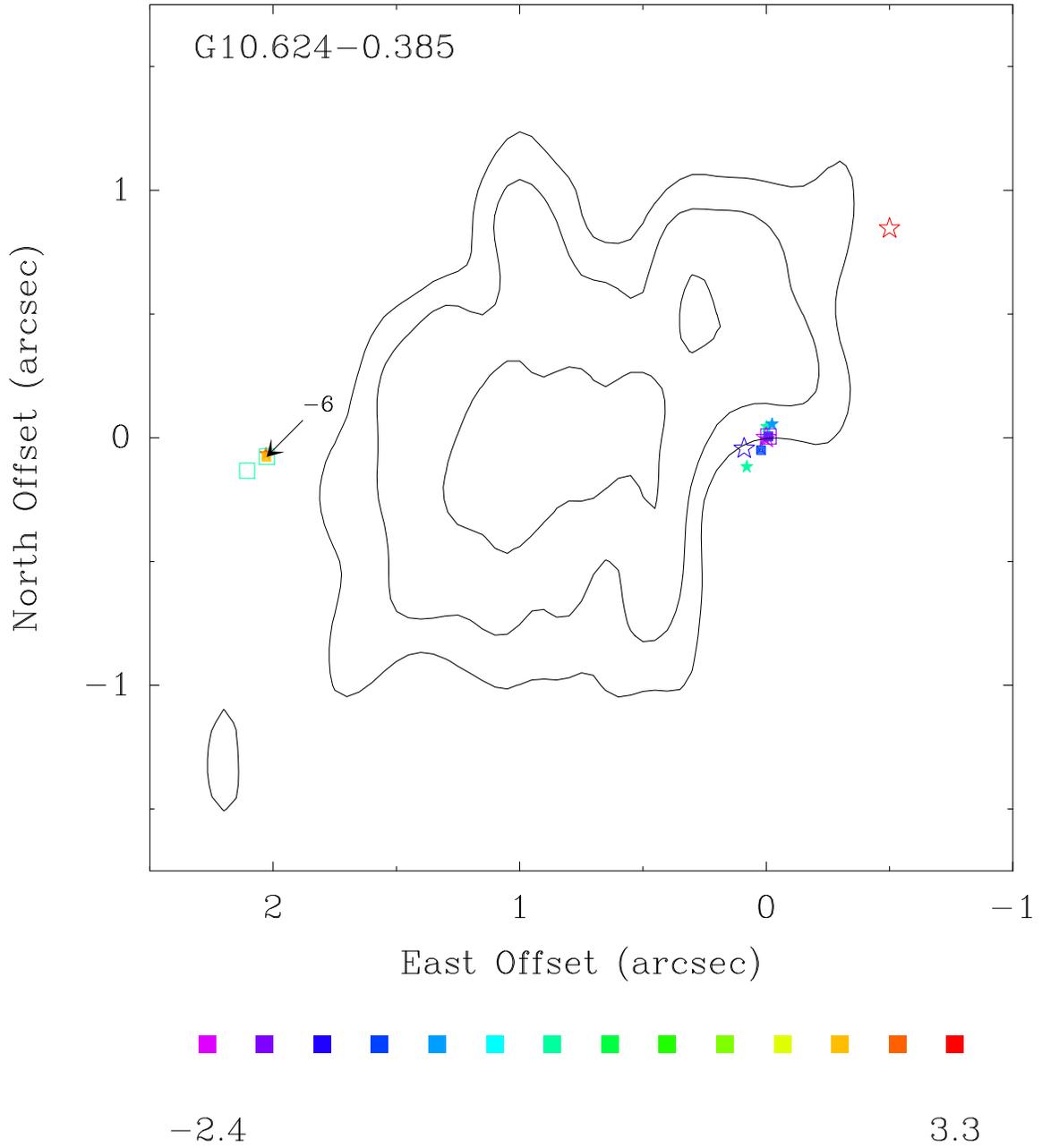}
\end{center}
\singlespace
\caption[Plot of maser spots in G10.624$-$0.385.]{Plot of maser spots
in G10.624$-$0.385.  Symbols are as in Figure \ref{g5v}.\label{g10v}}
\doublespace
\end{figure}

\begin{figure}
\begin{center}
\includegraphics[width=6.0in]{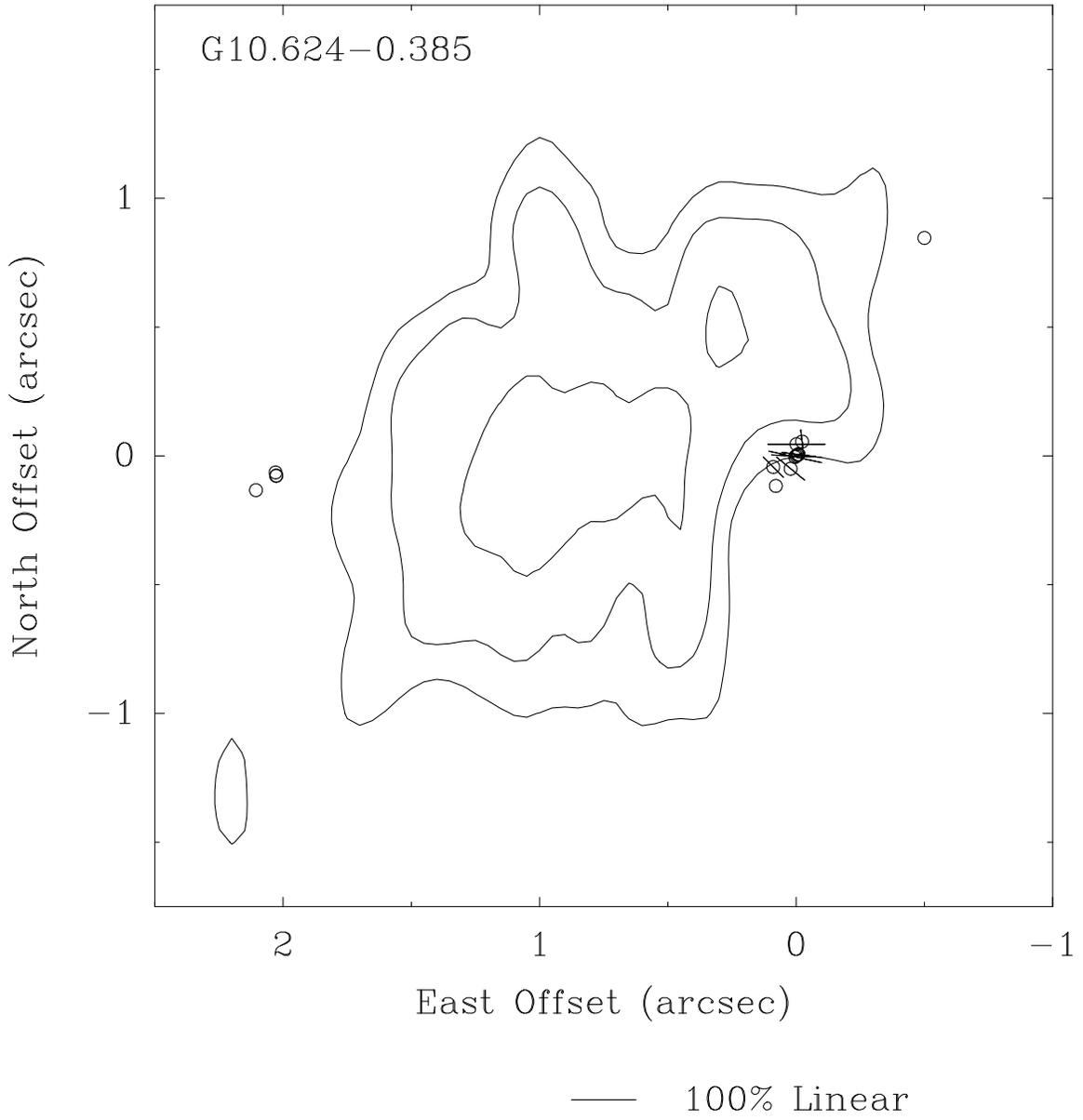}
\end{center}
\singlespace
\caption[Polarization map of G10.624$-$0.385.]{Polarization map of
G10.624$-$0.385.  Symbols are as in Figure \ref{g5p}.\label{g10p}}
\doublespace
\end{figure}

\begin{figure}
\begin{center}
\includegraphics[width=6.0in]{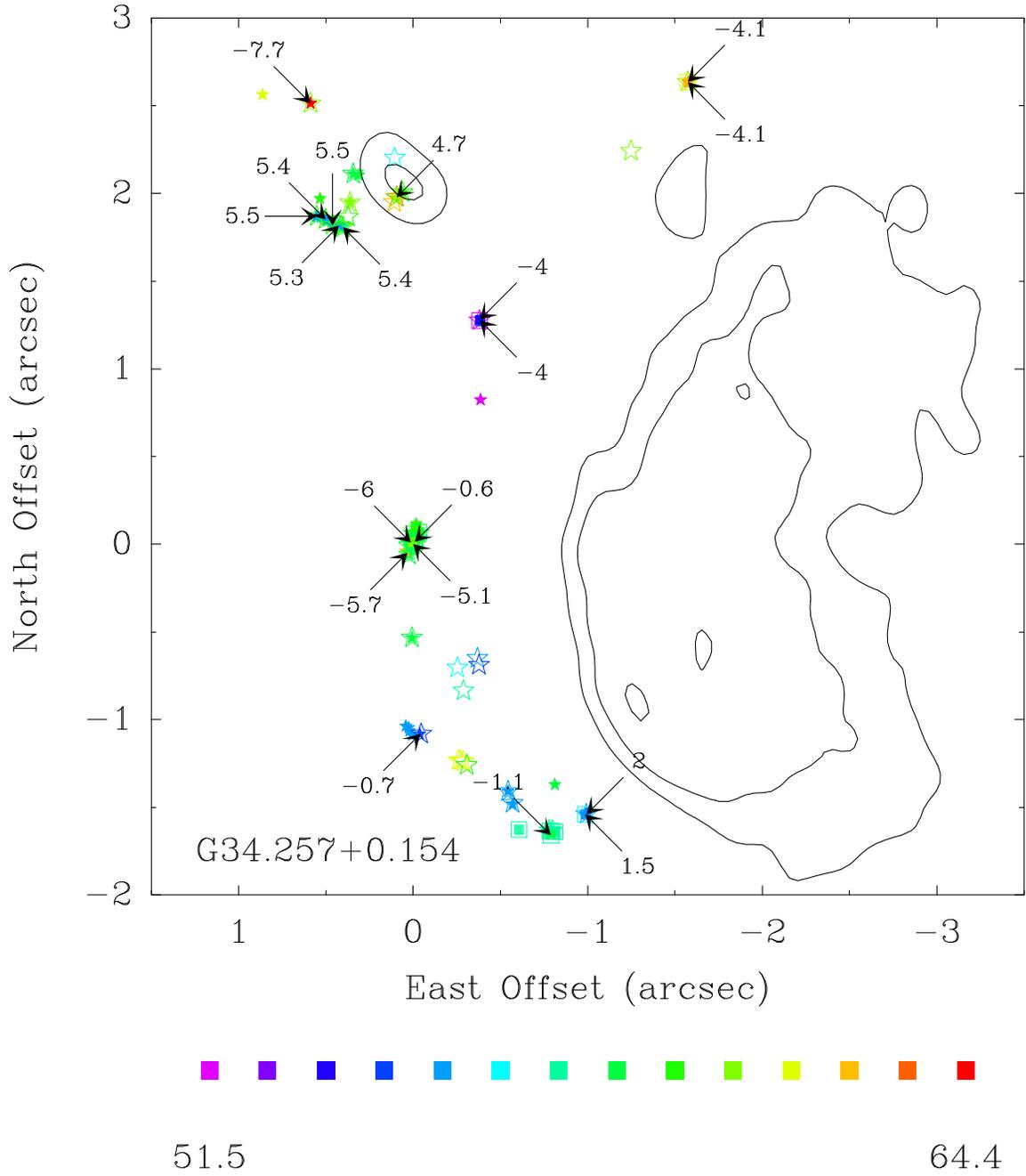}
\end{center}
\singlespace
\caption[Plot of maser spots in G34.257$+$0.154.]{Plot of maser spots
in G34.257$+$0.154.  Symbols are as in Figure \ref{g5v}.\label{g34v}}
\doublespace
\end{figure}

\begin{figure}
\begin{center}
\includegraphics[width=6.0in]{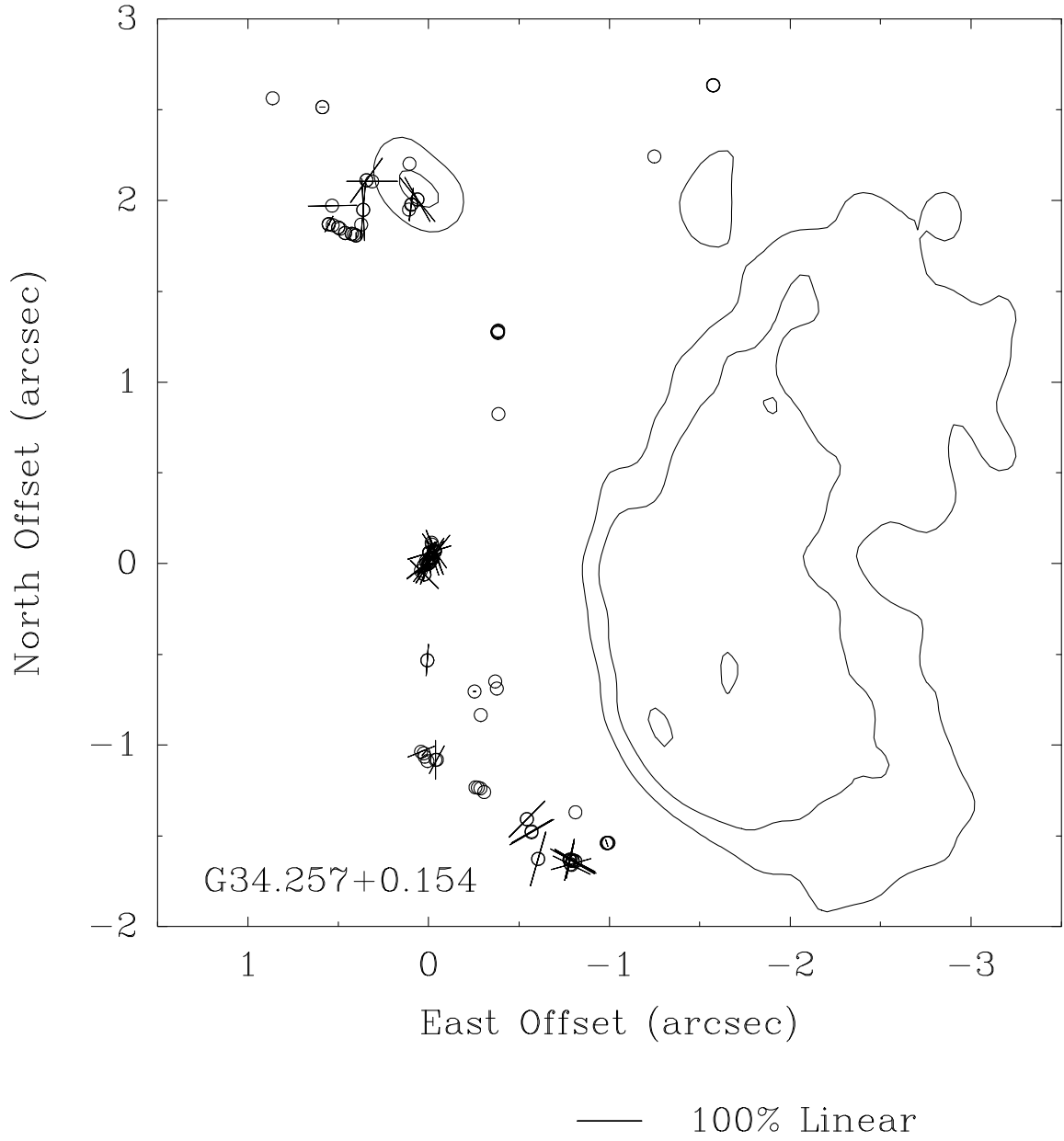}
\end{center}
\singlespace
\caption[Polarization map of G34.257$+$0.154.]{Polarization map of
G34.257$+$0.154.  Symbols are as in Figure \ref{g5p}.\label{g34p}}
\doublespace
\end{figure}

\begin{figure}
\begin{center}
\includegraphics[width=6.0in]{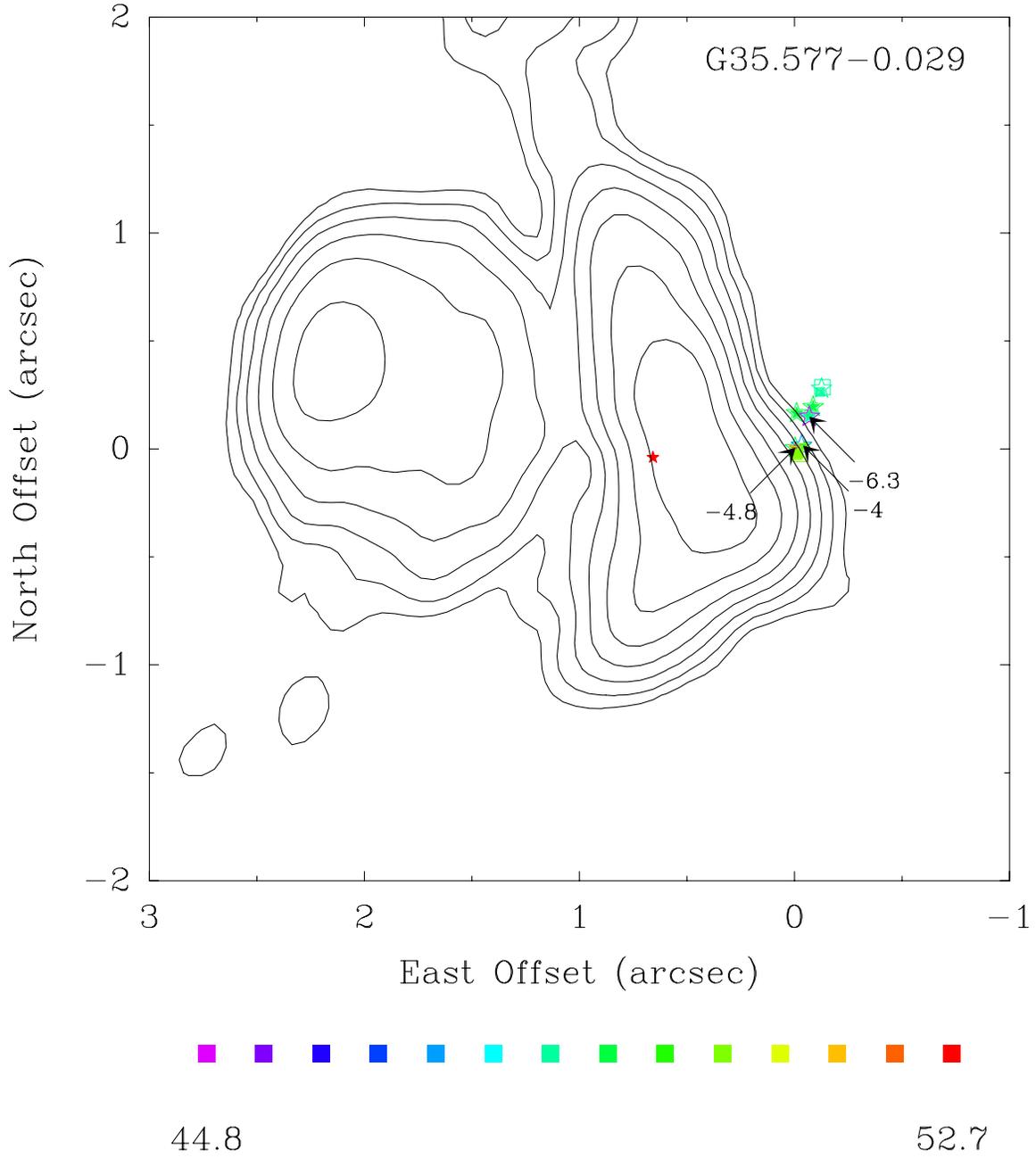}
\end{center}
\singlespace
\caption[Plot of maser spots in G35.577$-$0.029.]{Plot of maser spots
in G35.577$-$0.029.  Symbols are as in Figure \ref{g5v}.\label{g35v}}
\doublespace
\end{figure}

\begin{figure}
\begin{center}
\includegraphics[width=6.0in]{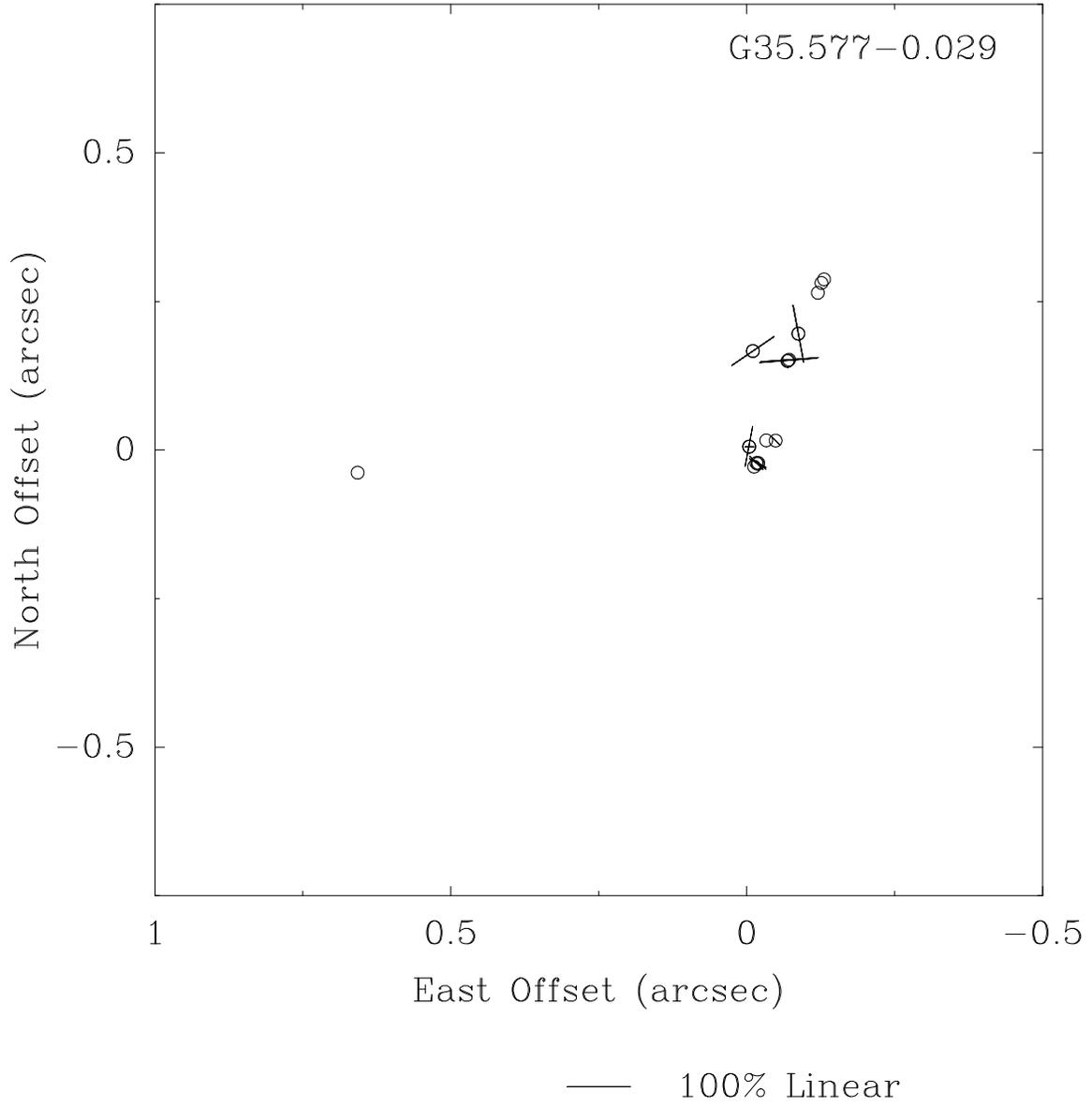}
\end{center}
\singlespace
\caption[Polarization map of G35.577$-$0.029.]{Polarization map of
G35.577$-$0.029.  Symbols are as in Figure \ref{g5p}.  The continuum
image is suppressed for clarity.\label{g35p}}
\doublespace
\end{figure}

\begin{figure}
\begin{center}
\includegraphics[width=6.0in]{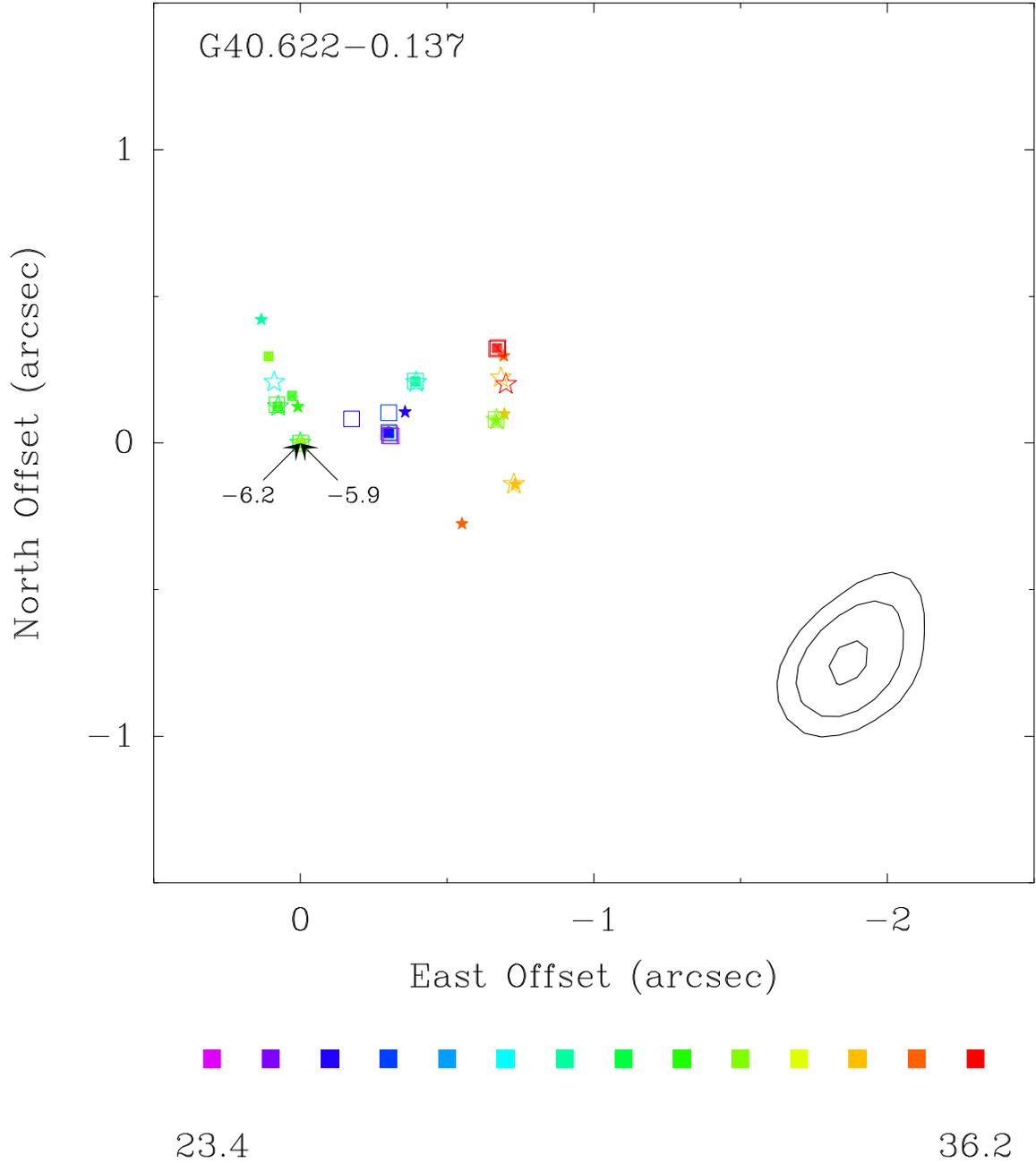}
\end{center}
\singlespace
\caption[Plot of maser spots in G40.622$-$0.137.]{Plot of maser spots
in G40.622$-$0.137.  Symbols are as in Figure \ref{g5v}.\label{g40v}}
\doublespace
\end{figure}

\begin{figure}
\begin{center}
\includegraphics[width=6.0in]{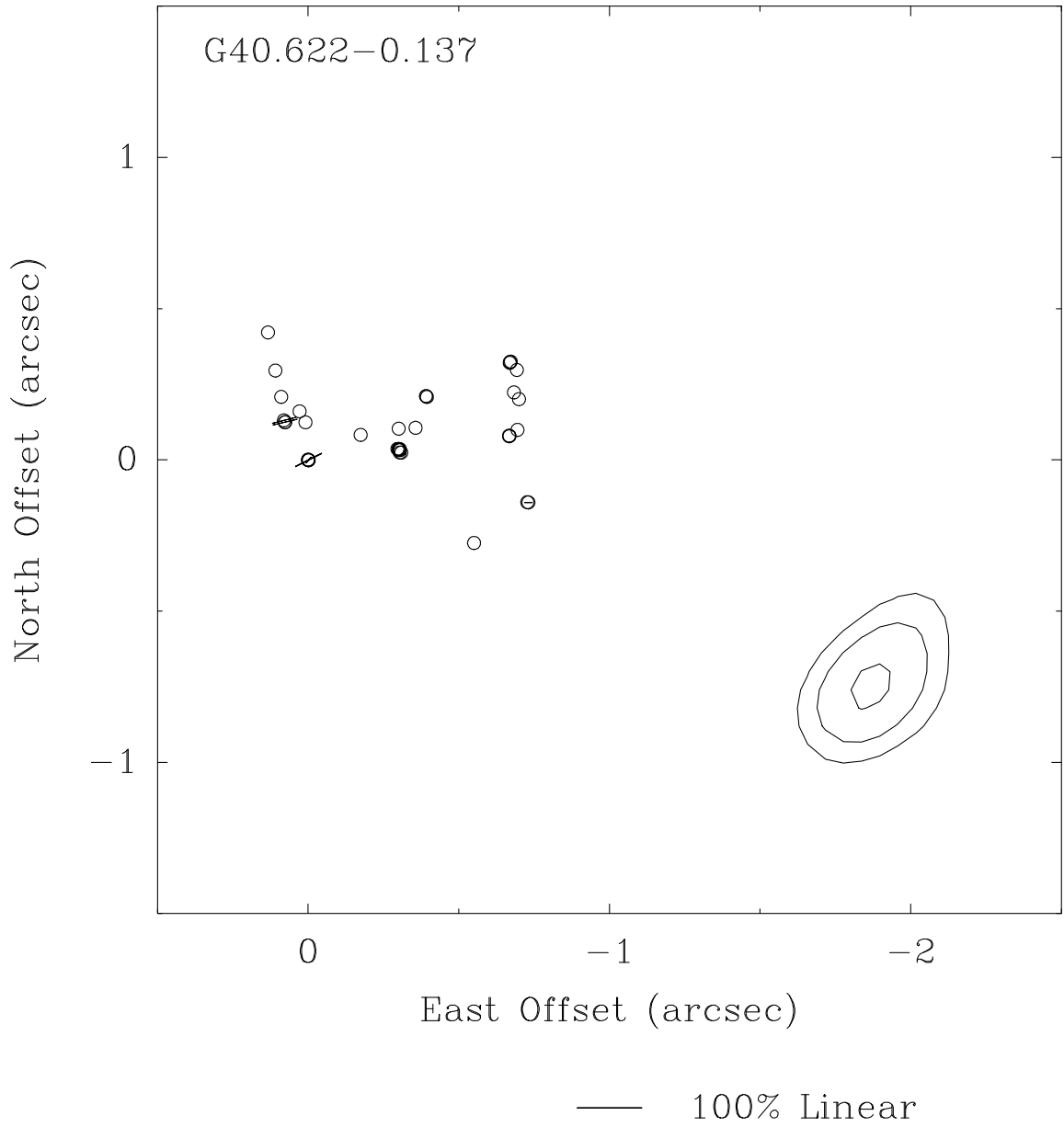}
\end{center}
\singlespace
\caption[Polarization map of G40.622$-$0.137.]{Polarization map of
G40.622$-$0.137.  Symbols are as in Figure \ref{g5p}.\label{g40p}}
\doublespace
\end{figure}

\begin{figure}
\begin{center}
\includegraphics[width=6.0in]{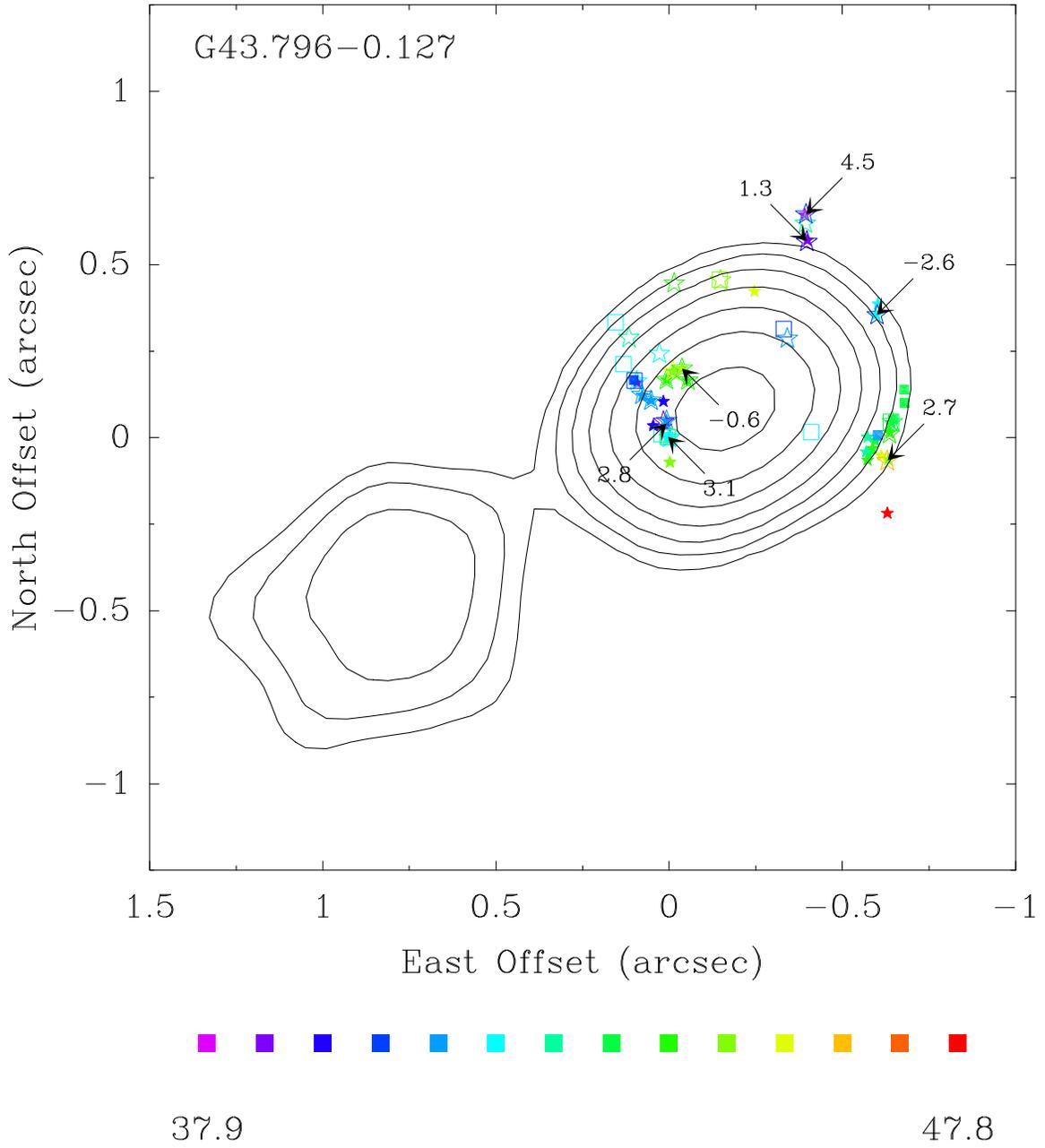}
\end{center}
\singlespace
\caption[Plot of maser spots in G43.796$-$0.127.]{Plot of maser spots
in G43.796$-$0.127.  Symbols are as in Figure \ref{g5v}.\label{g43v}}
\doublespace
\end{figure}

\begin{figure}
\begin{center}
\includegraphics[width=6.0in]{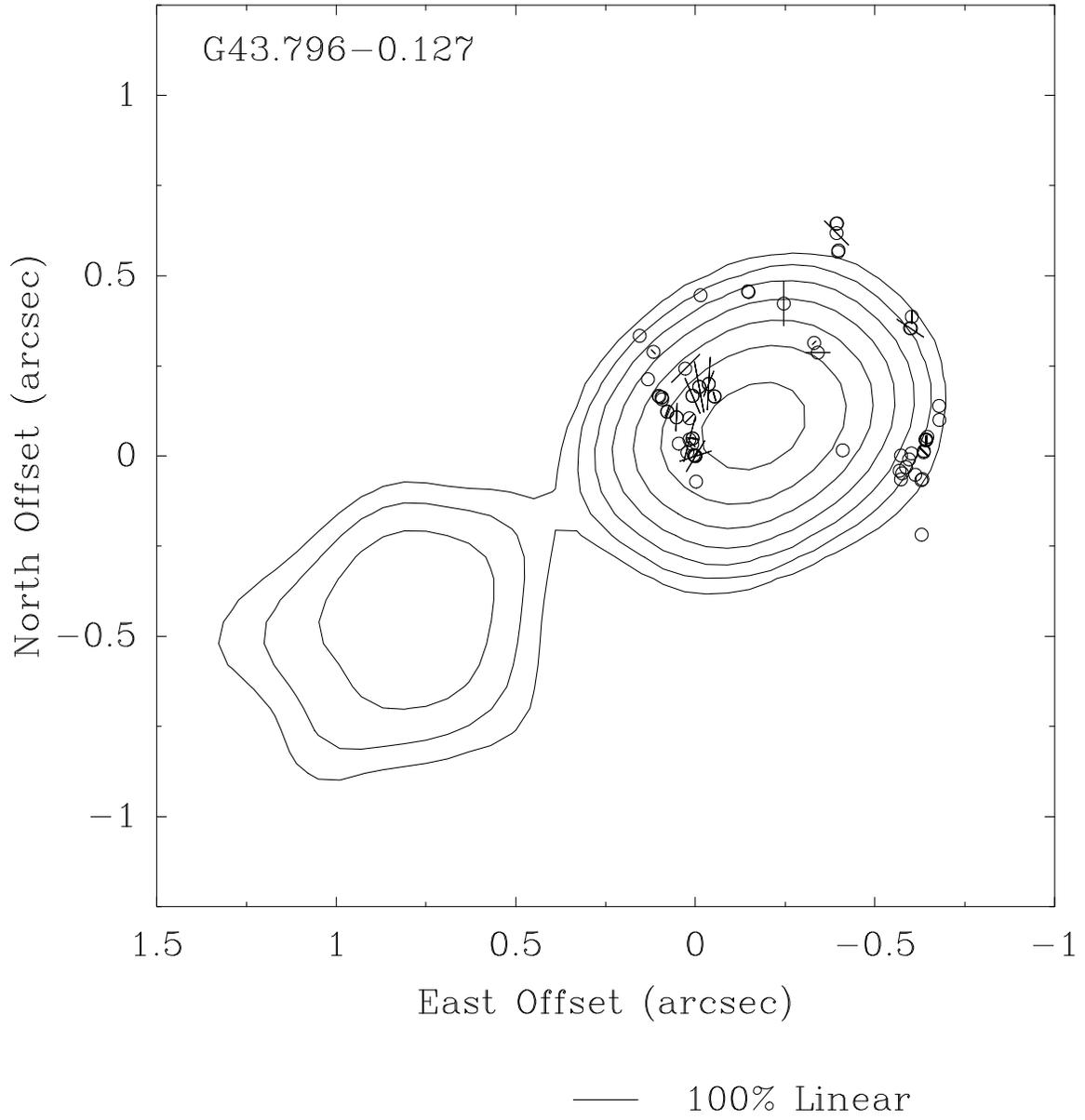}
\end{center}
\singlespace
\caption[Polarization map of G43.796$-$0.127.]{Polarization map of
G43.796$-$0.127.  Symbols are as in Figure \ref{g5p}.\label{g43p}}
\doublespace
\end{figure}

\begin{figure}
\begin{center}
\includegraphics[width=6.0in]{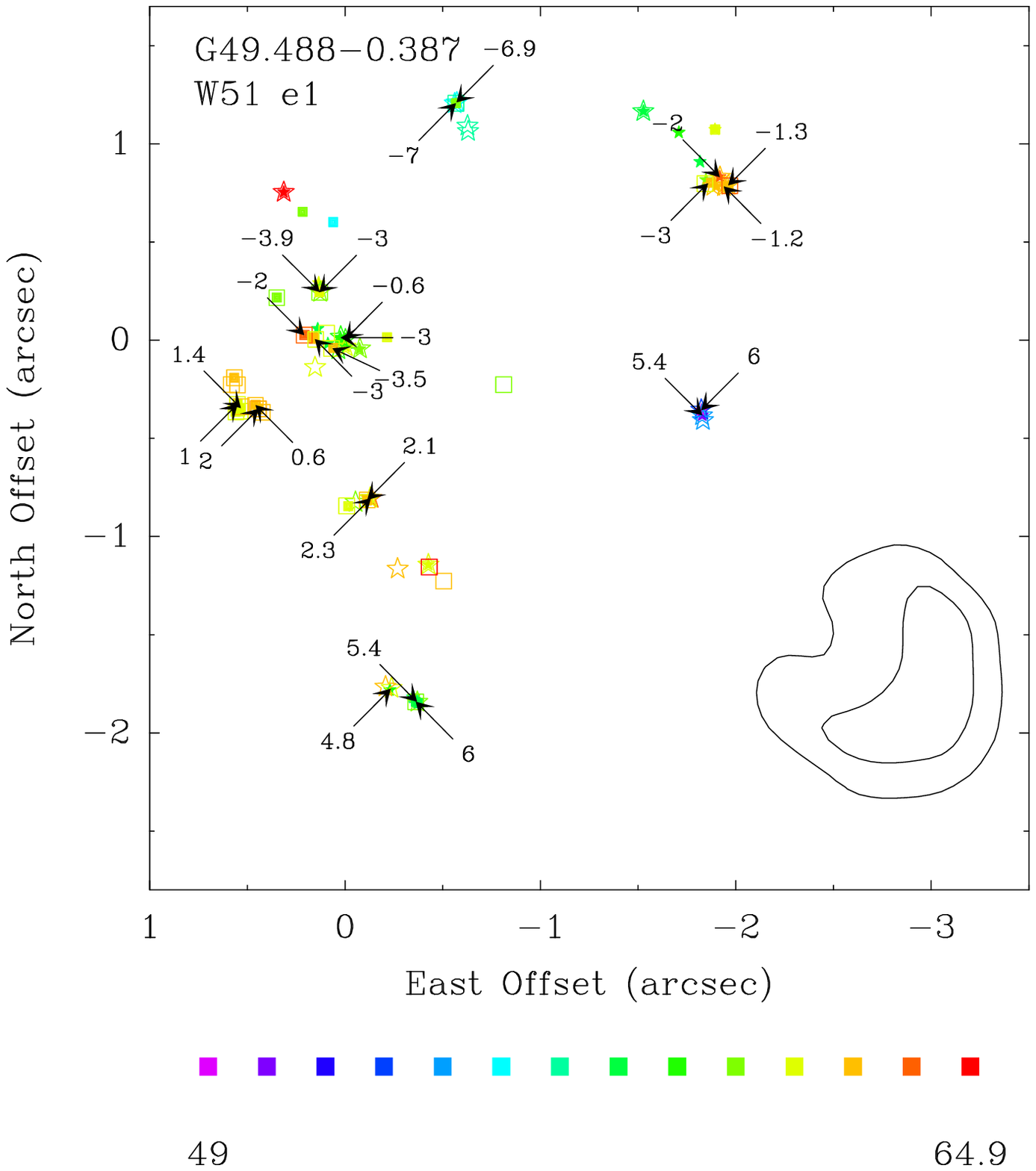}
\end{center}
\singlespace
\caption[Plot of maser spots in W51 M/S e1.]{Plot of maser spots in
W51 M/S e1.  Symbols are as in Figure \ref{g5v}.\label{w51e1v}}
\doublespace
\end{figure}

\begin{figure}
\begin{center}
\includegraphics[width=6.0in]{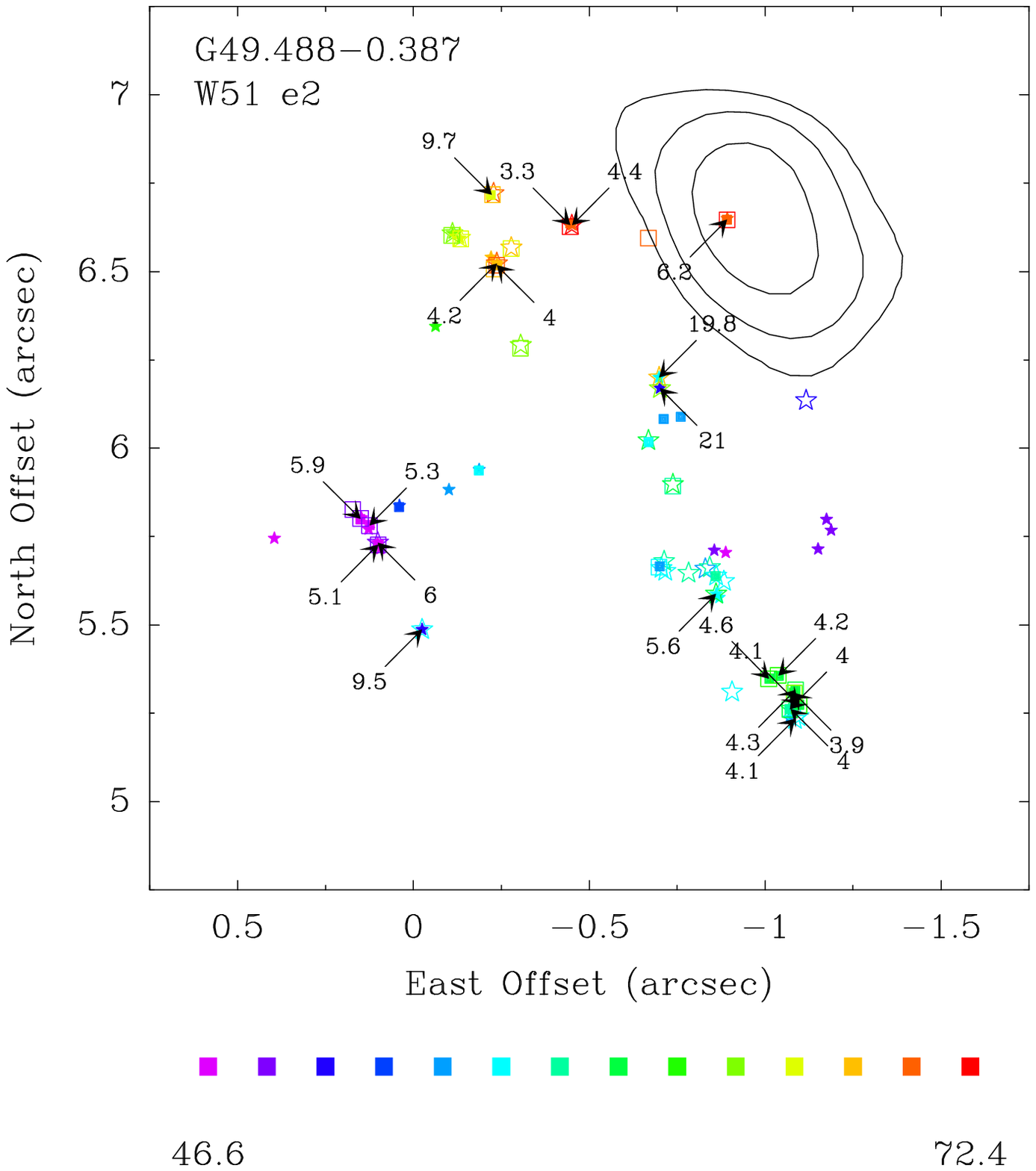}
\end{center}
\singlespace
\caption[Plot of maser spots in W51 M/S e2.]{Plot of maser spots in
W51 M/S e2.  Symbols are as in Figure \ref{g5v}.  Coordinates are
offsets from the origin in Figure \ref{w51e1v}.\label{w51e2v}}
\doublespace
\end{figure}

\begin{figure}
\begin{center}
\includegraphics[width=6.0in]{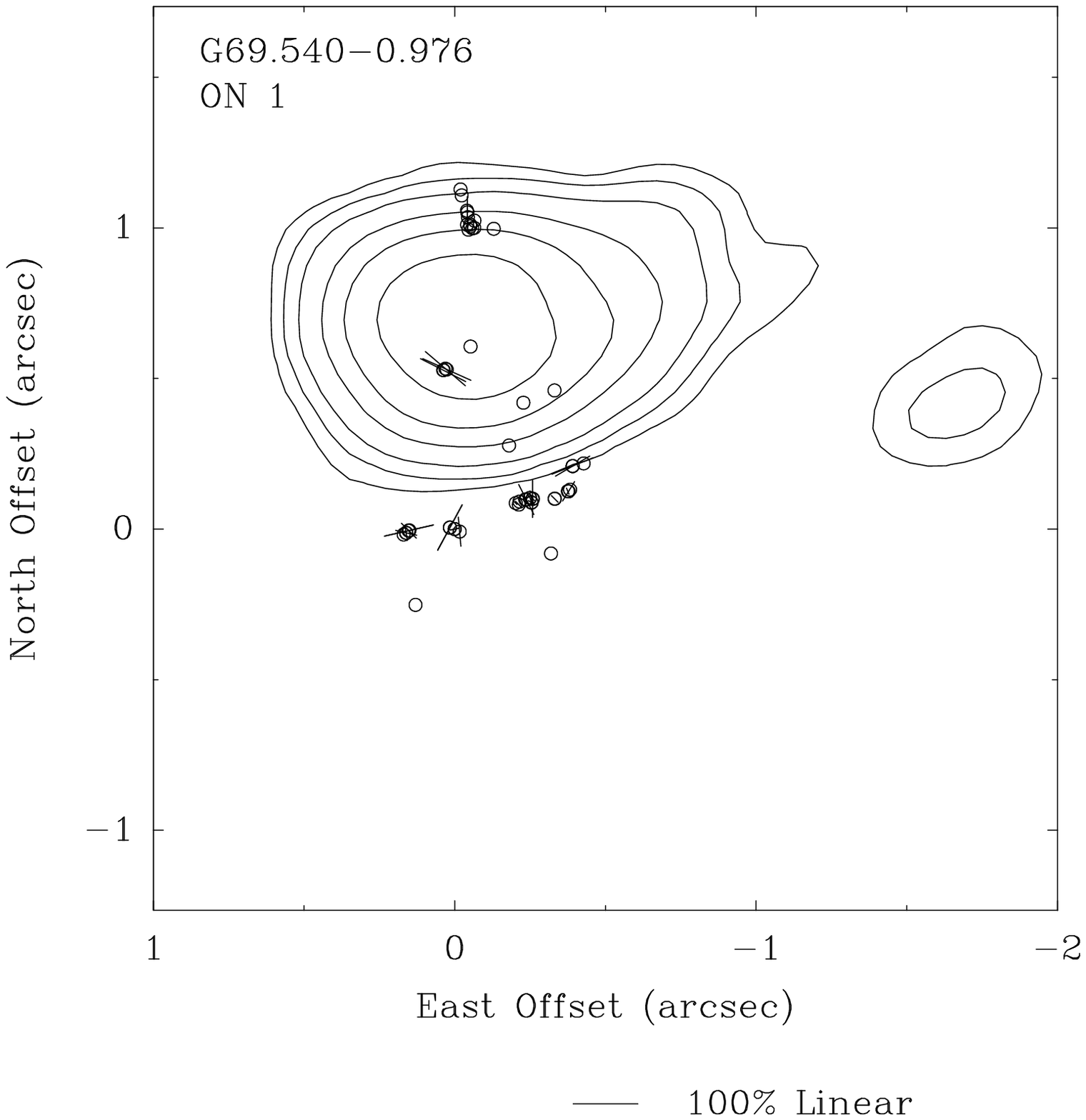}
\end{center}
\singlespace
\caption[Polarization map of ON 1.]{Polarization map of ON 1.  Symbols
are as in Figure \ref{g5p}.\label{on1p}}
\doublespace
\end{figure}

\begin{figure}
\begin{center}
\includegraphics[width=6.0in]{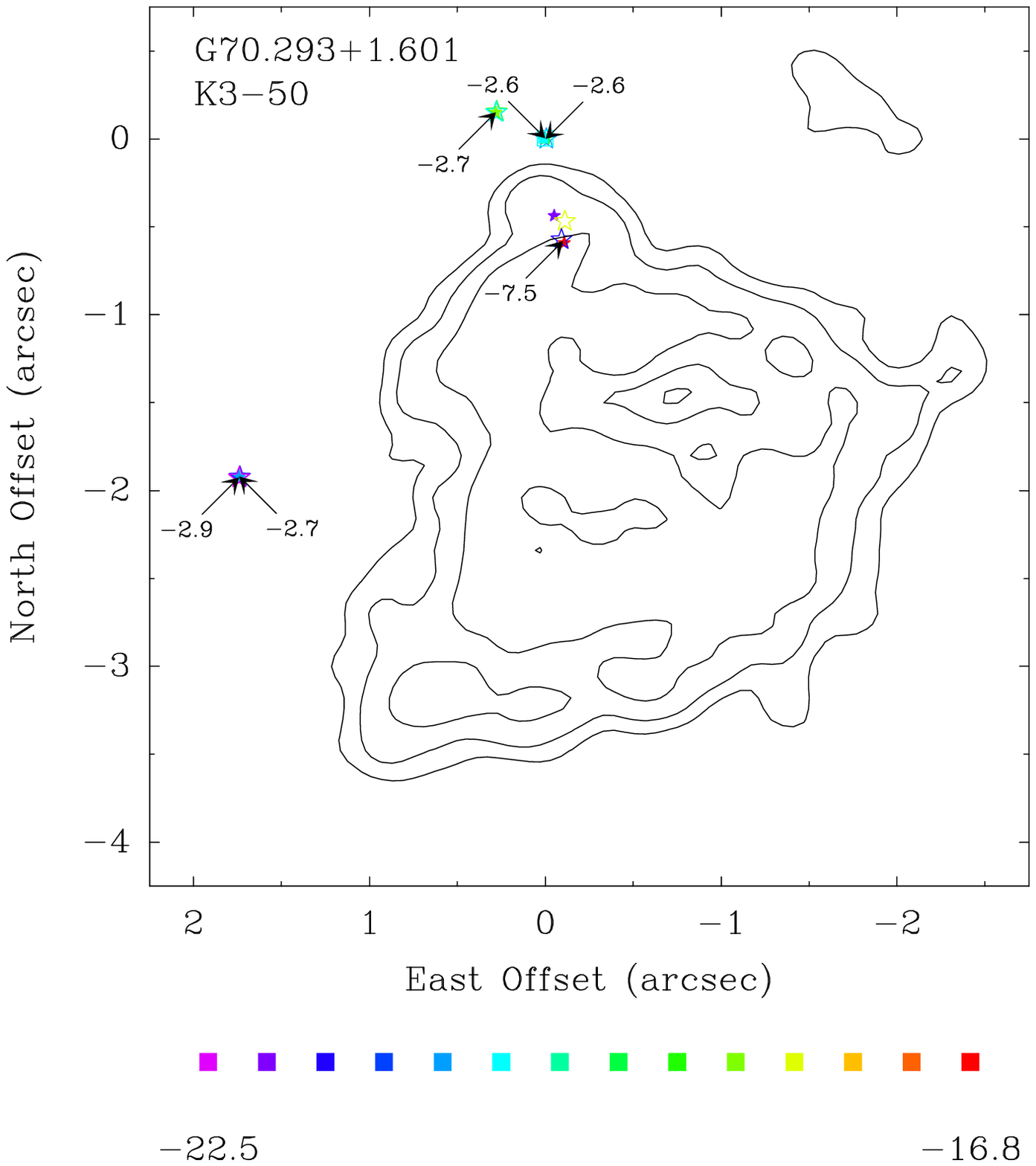}
\end{center}
\singlespace
\caption[Plot of maser spots in K3$-$50.]{Plot of maser spots in
K3$-$50.  Symbols are as in Figure \ref{g5v}.\label{k350v}}
\doublespace
\end{figure}

\begin{figure}
\begin{center}
\includegraphics[width=6.0in]{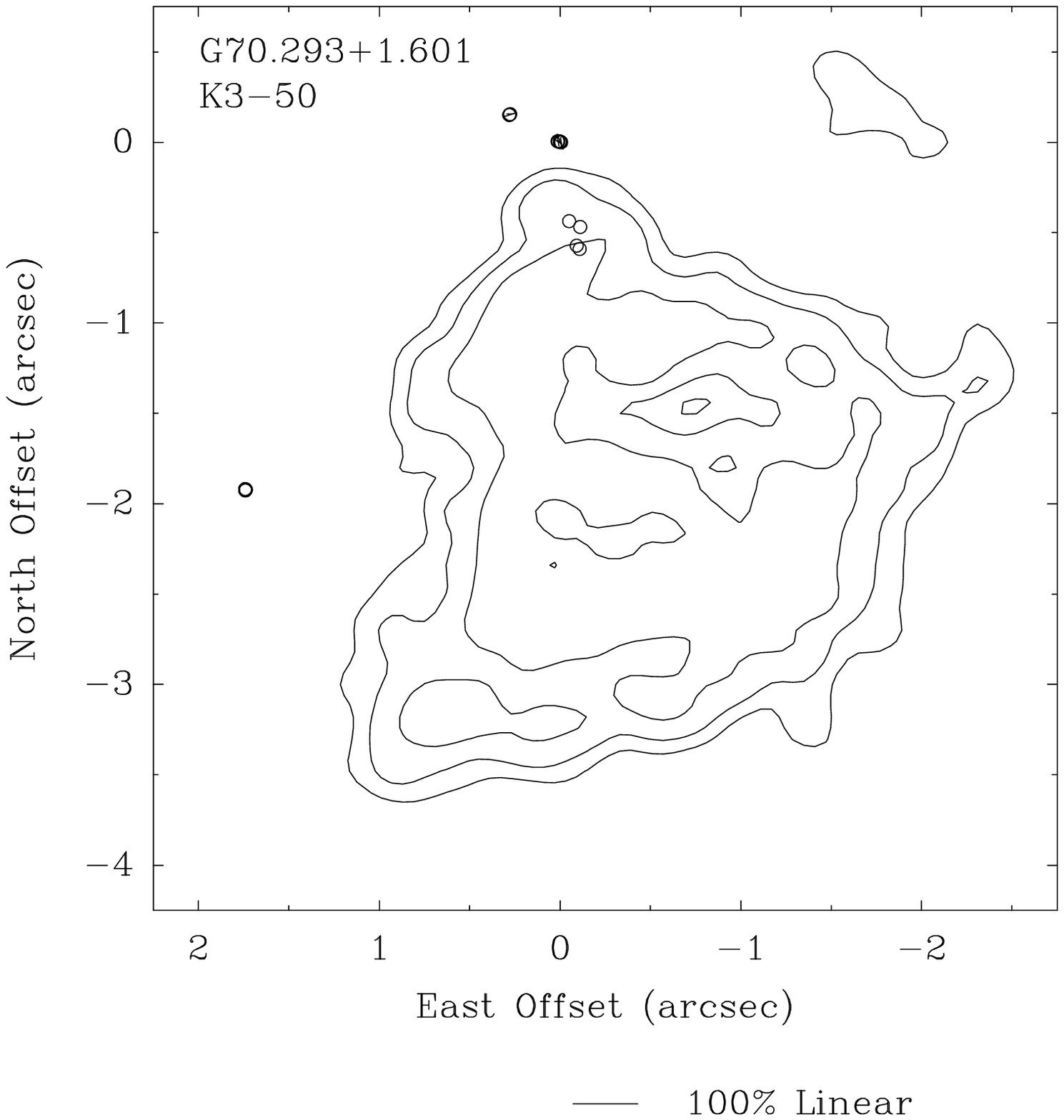}
\end{center}
\singlespace
\caption[Polarization map of K3$-$50.]{Polarization map of K3$-$50.
Symbols are as in Figure \ref{g5p}.\label{k350p}}
\doublespace
\end{figure}

\begin{figure}
\begin{center}
\includegraphics[width=6.0in]{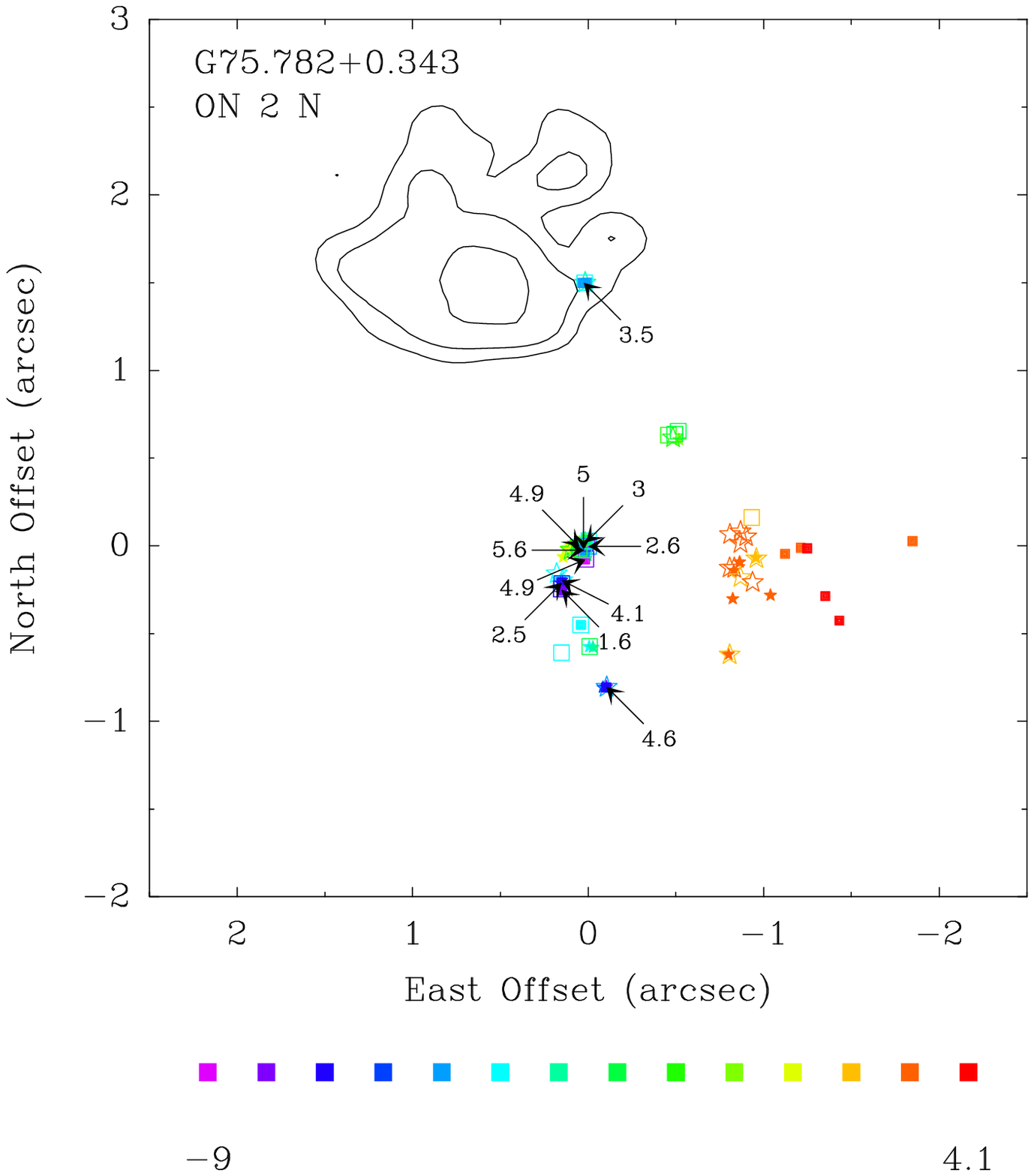}
\end{center}
\singlespace
\caption[Plot of maser spots in ON 2 N.]{Plot of maser spots in ON 2
N.  Symbols are as in Figure \ref{g5v}.\label{on2nv}}
\doublespace
\end{figure}

\begin{figure}
\begin{center}
\includegraphics[width=6.0in]{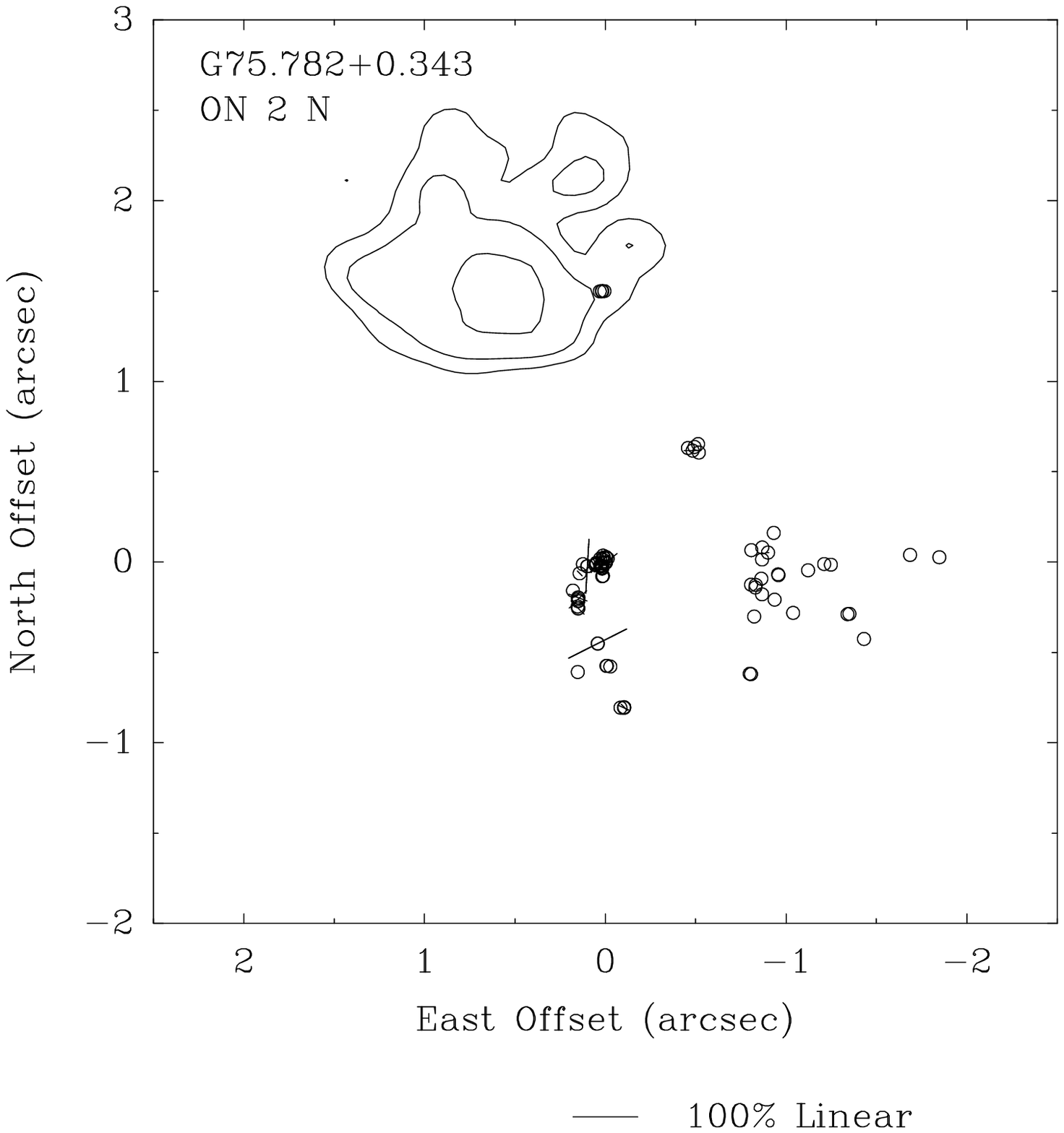}
\end{center}
\singlespace
\caption[Polarization map of ON 2 N.]{Polarization map of ON 2 N.
Symbols are as in Figure \ref{g5p}.\label{on2np}}
\doublespace
\end{figure}

\begin{figure}
\begin{center}
\includegraphics[width=6.0in]{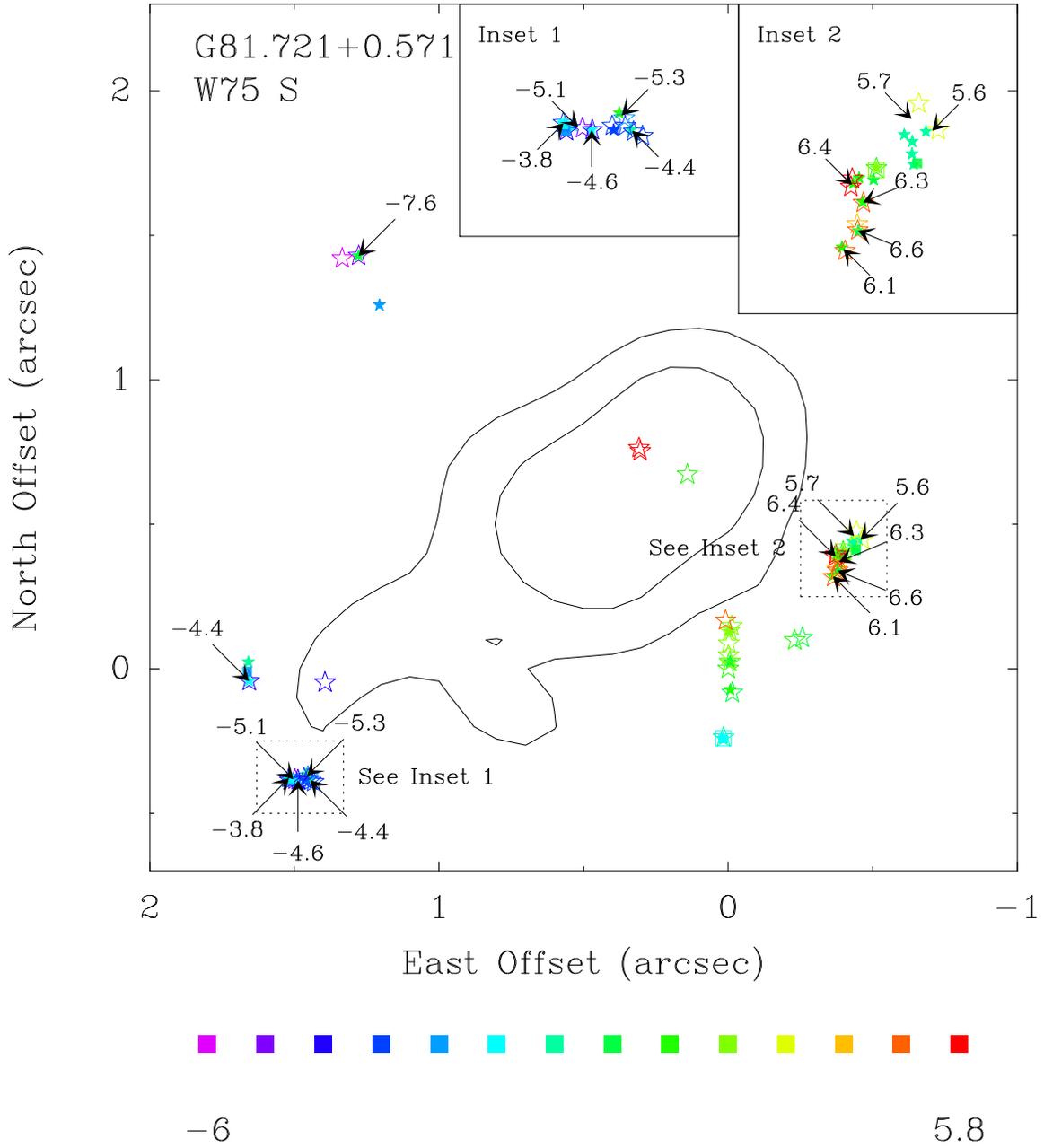}
\end{center}
\singlespace
\caption[Plot of maser spots in W75 S.]{Plot of maser spots in W75 S.
  The masers are superposed on a U-band continuum map.
Symbols are as in Figure \ref{g5v}.\label{w75sv}}
\doublespace
\end{figure}

\begin{figure}
\begin{center}
\includegraphics[width=6.0in]{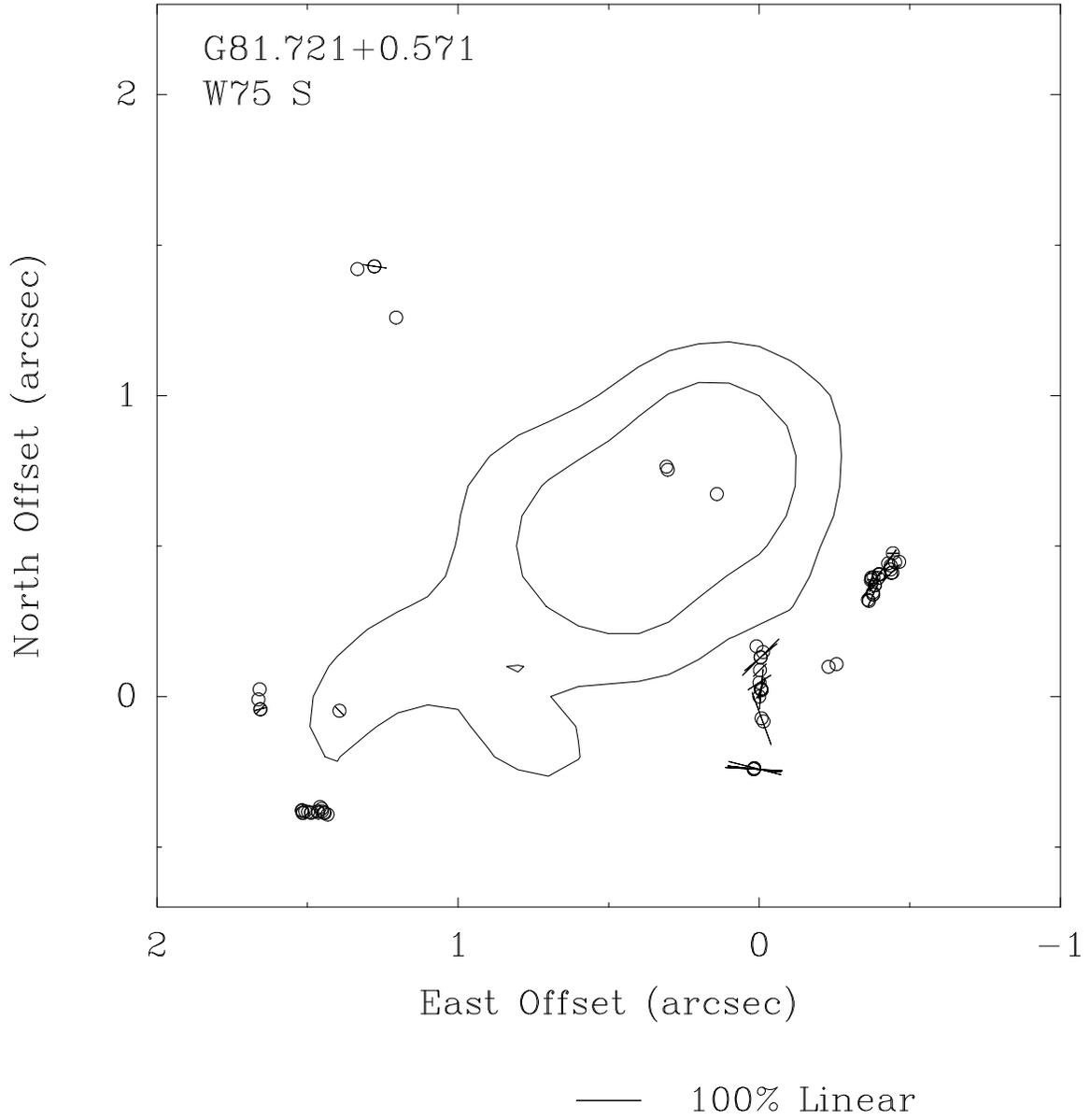}
\end{center}
\singlespace
\caption[Polarization map of W75 S.]{Polarization map of W75 S.
  The masers are superposed on a U-band continuum map.
Symbols are as in Figure \ref{g5p}.\label{w75sp}}
\doublespace
\end{figure}

\begin{figure}
\begin{center}
\includegraphics[width=6.0in]{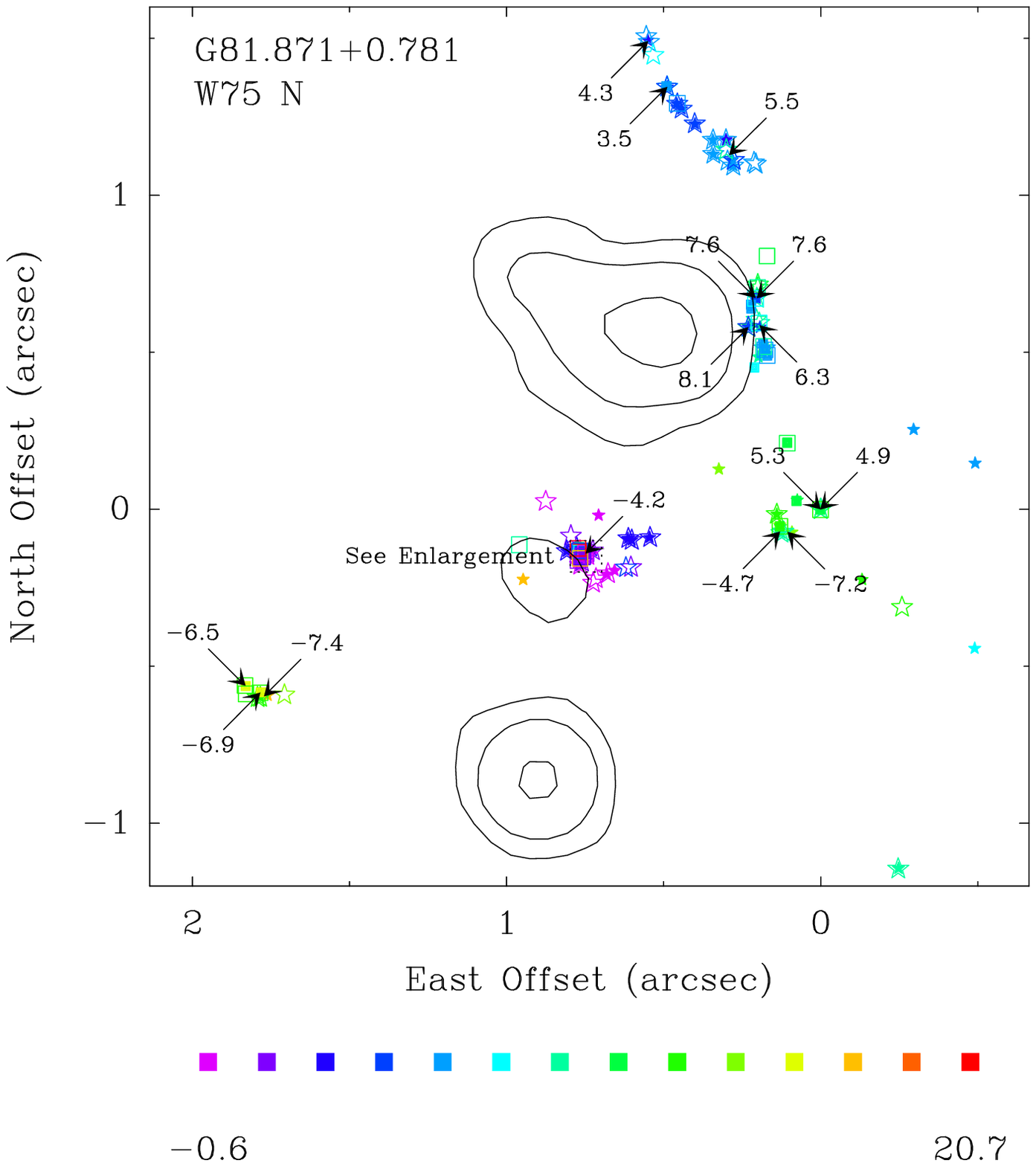}
\end{center}
\singlespace
\caption[Plot of maser spots in W75 N.]{Plot of maser spots in W75 N.
Symbols are as in Figure \ref{g5v}.\label{w75nv}}
\doublespace
\end{figure}

\begin{figure}
\begin{center}
\includegraphics[width=6.0in]{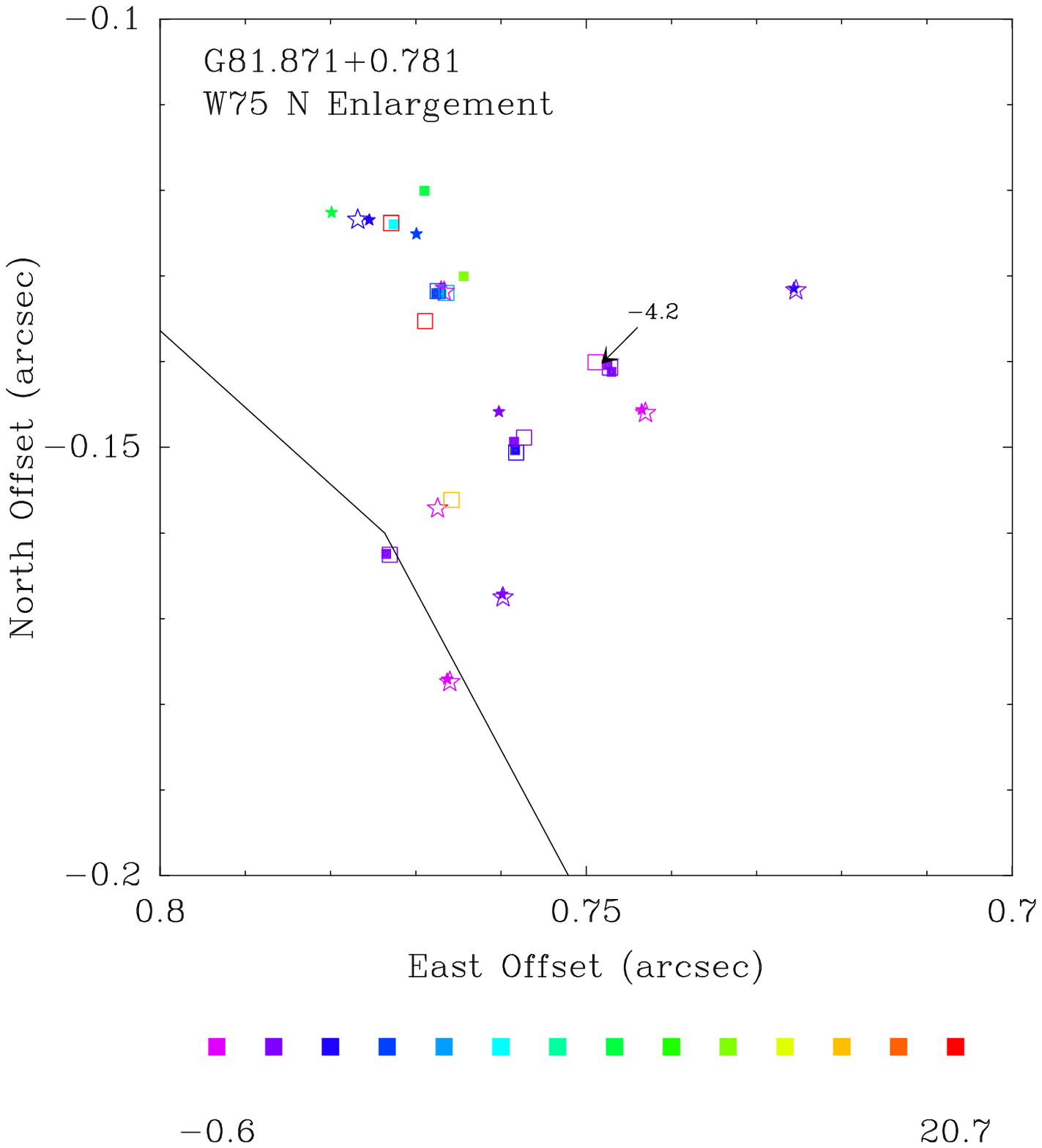}
\end{center}
\singlespace
\caption[Enlargement of cluster of maser spots in W75 N.]{Enlargement
  of cluster of maser spots in W75 N in Figure \ref{w75nv}.  Symbols
  are as in Figure \ref{g5v}.\label{w75nvb}}
\doublespace
\end{figure}

\begin{figure}
\begin{center}
\includegraphics[width=6.0in]{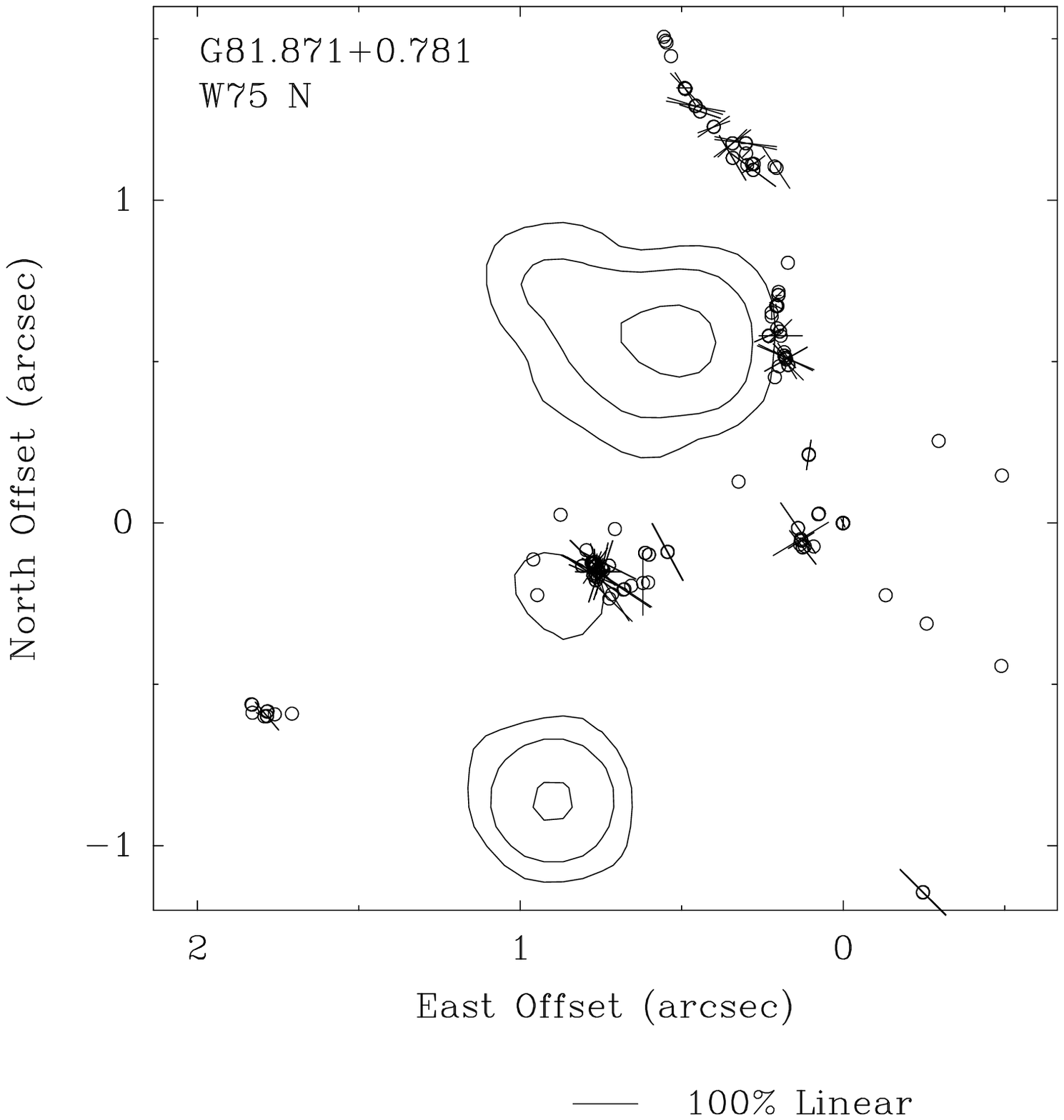}
\end{center}
\singlespace
\caption[Polarization map of W75 N.]{Polarization map of W75 N.
Symbols are as in Figure \ref{g5p}.\label{w75np}}
\doublespace
\end{figure}

\begin{figure}
\begin{center}
\includegraphics[width=6.0in]{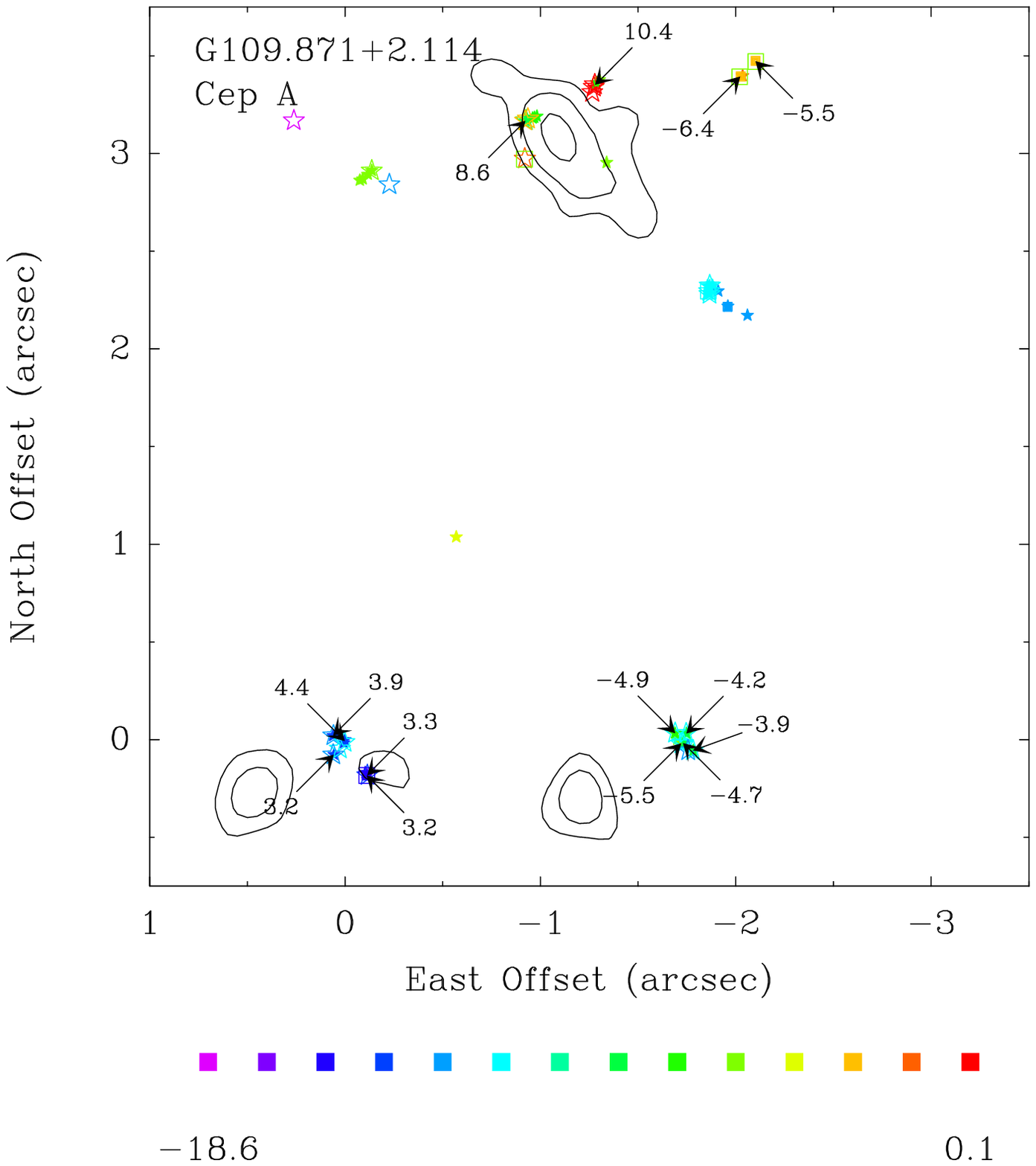}
\end{center}
\singlespace
\caption[Plot of maser spots in Cep A.]{Plot of maser spots in Cep A.
Symbols are as in Figure \ref{g5v}.\label{cepav}}
\doublespace
\end{figure}

\begin{figure}
\begin{center}
\includegraphics[width=6.0in]{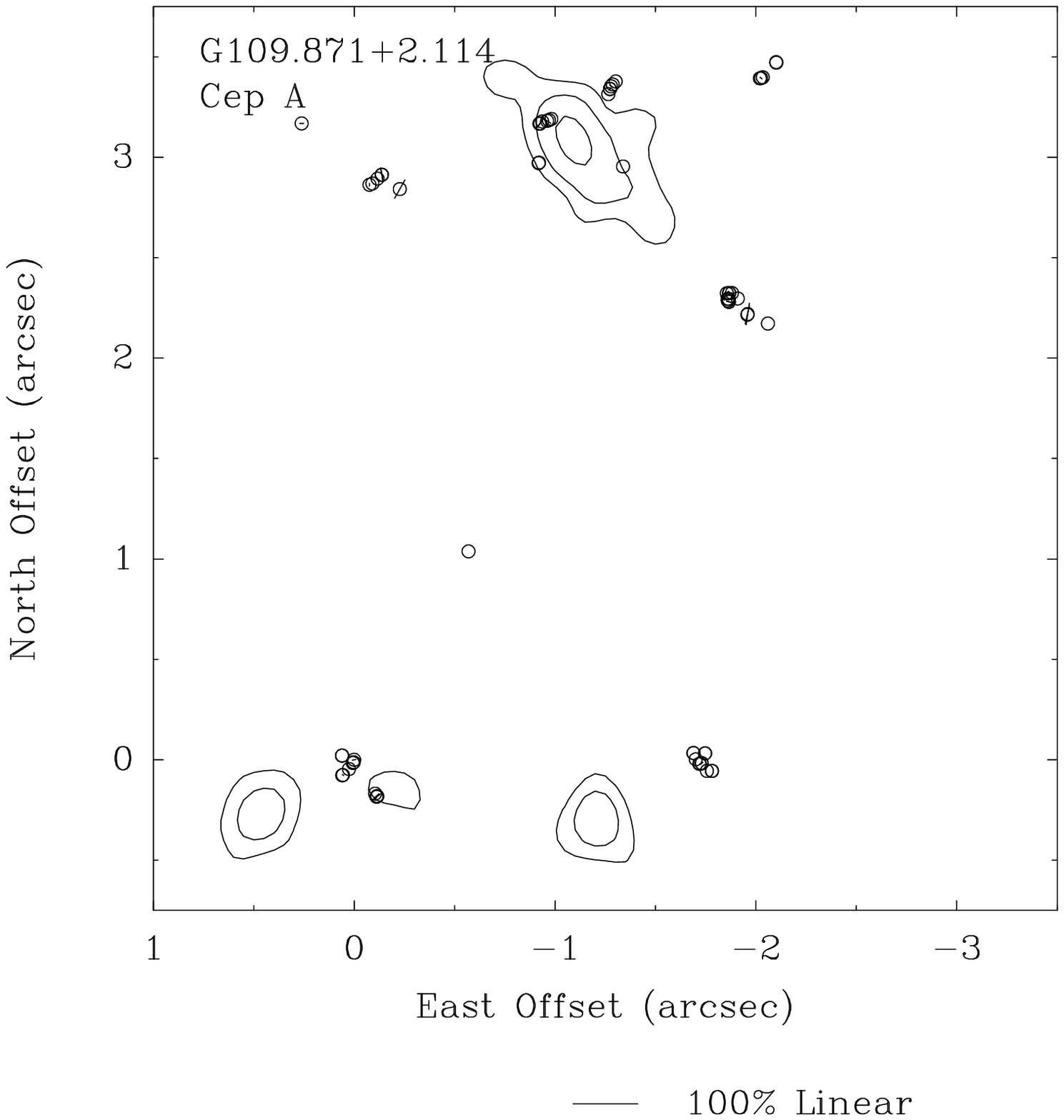}
\end{center}
\singlespace
\caption[Polarization map of Cep A.]{Polarization map of Cep A.
Symbols are as in Figure \ref{g5p}.\label{cepap}}
\doublespace
\end{figure}

\begin{figure}
\begin{center}
\includegraphics[width=6.0in]{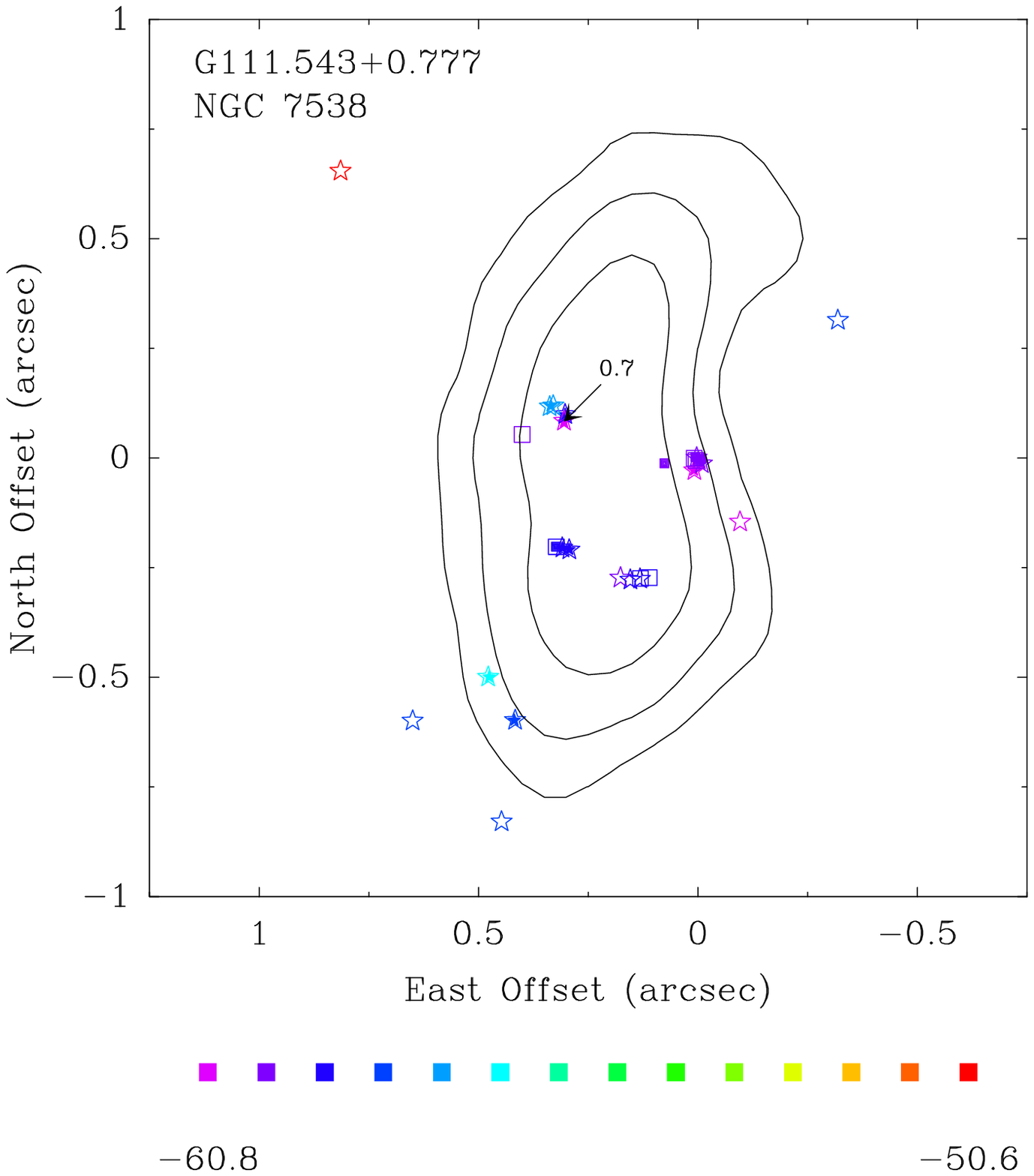}
\end{center}
\singlespace
\caption[Plot of maser spots in NGC 7538.]{Plot of maser spots in NGC
7538.  Symbols are as in Figure \ref{g5v}.\label{n7538v}}
\doublespace
\end{figure}

\begin{figure}
\begin{center}
\includegraphics[width=6.0in]{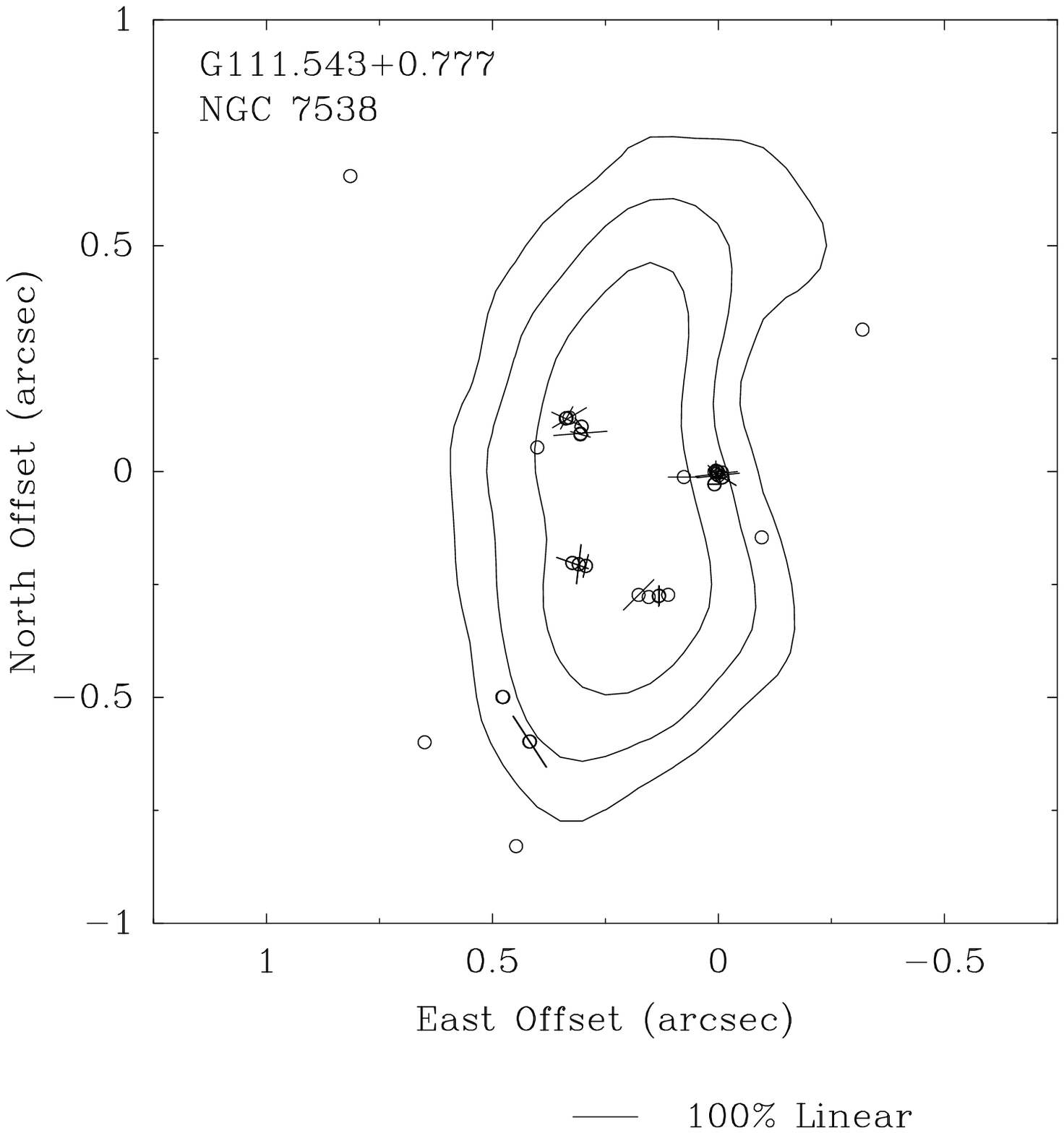}
\end{center}
\singlespace
\caption[Polarization map of NGC 7538.]{Polarization map of NGC 7538.
Symbols are as in Figure \ref{g5p}.\label{n7538p}}
\doublespace
\end{figure}

\begin{figure}
\begin{center}
\includegraphics[width=6.0in]{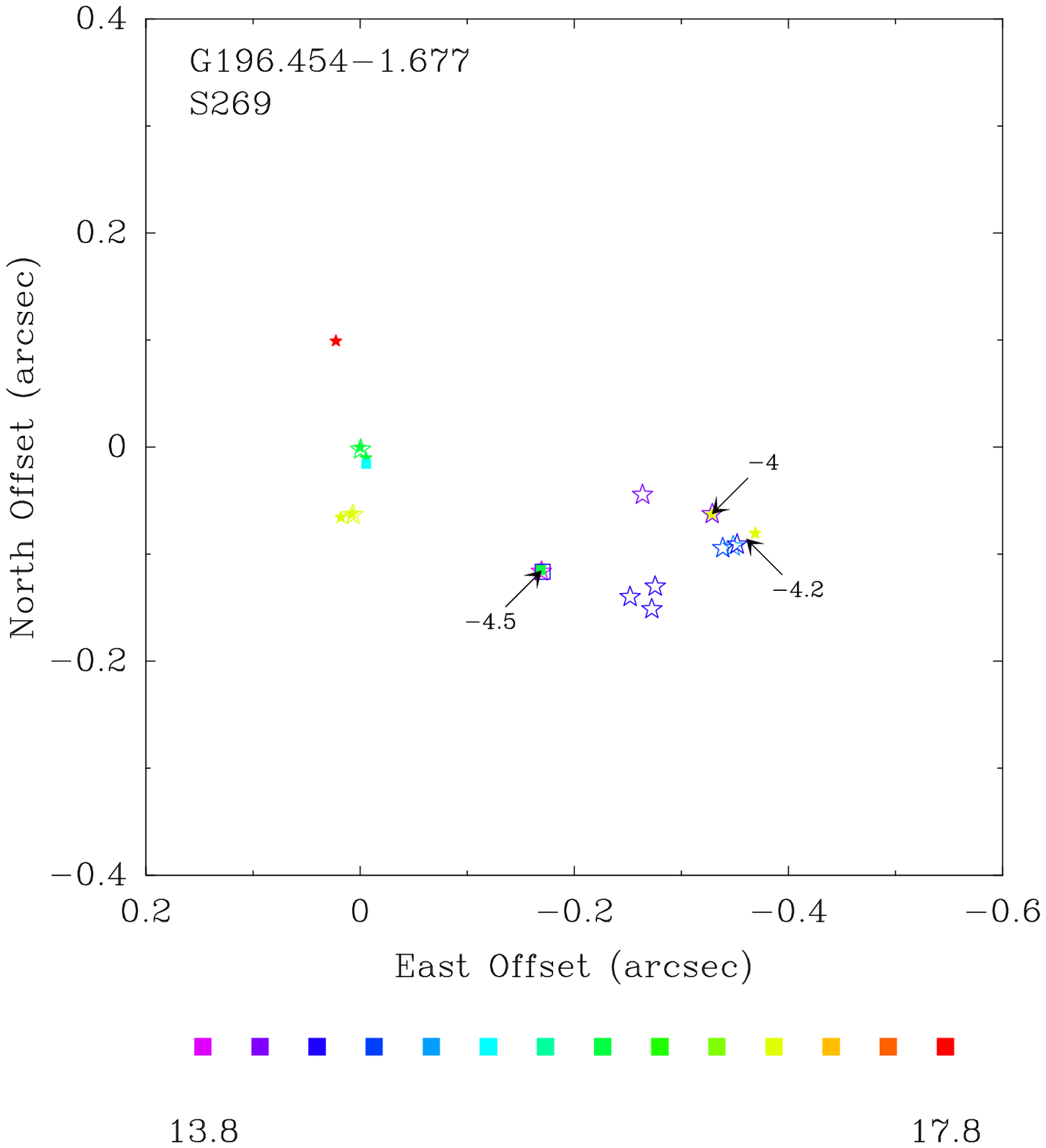}
\end{center}
\singlespace
\caption[Plot of maser spots in S269.]{Plot of maser spots in S269.
There is no detectable continuum emission in the region.  Symbols are
as in Figure \ref{g5v}.\label{s269v}}
\doublespace
\end{figure}

\begin{figure}
\begin{center}
\includegraphics[width=6.0in]{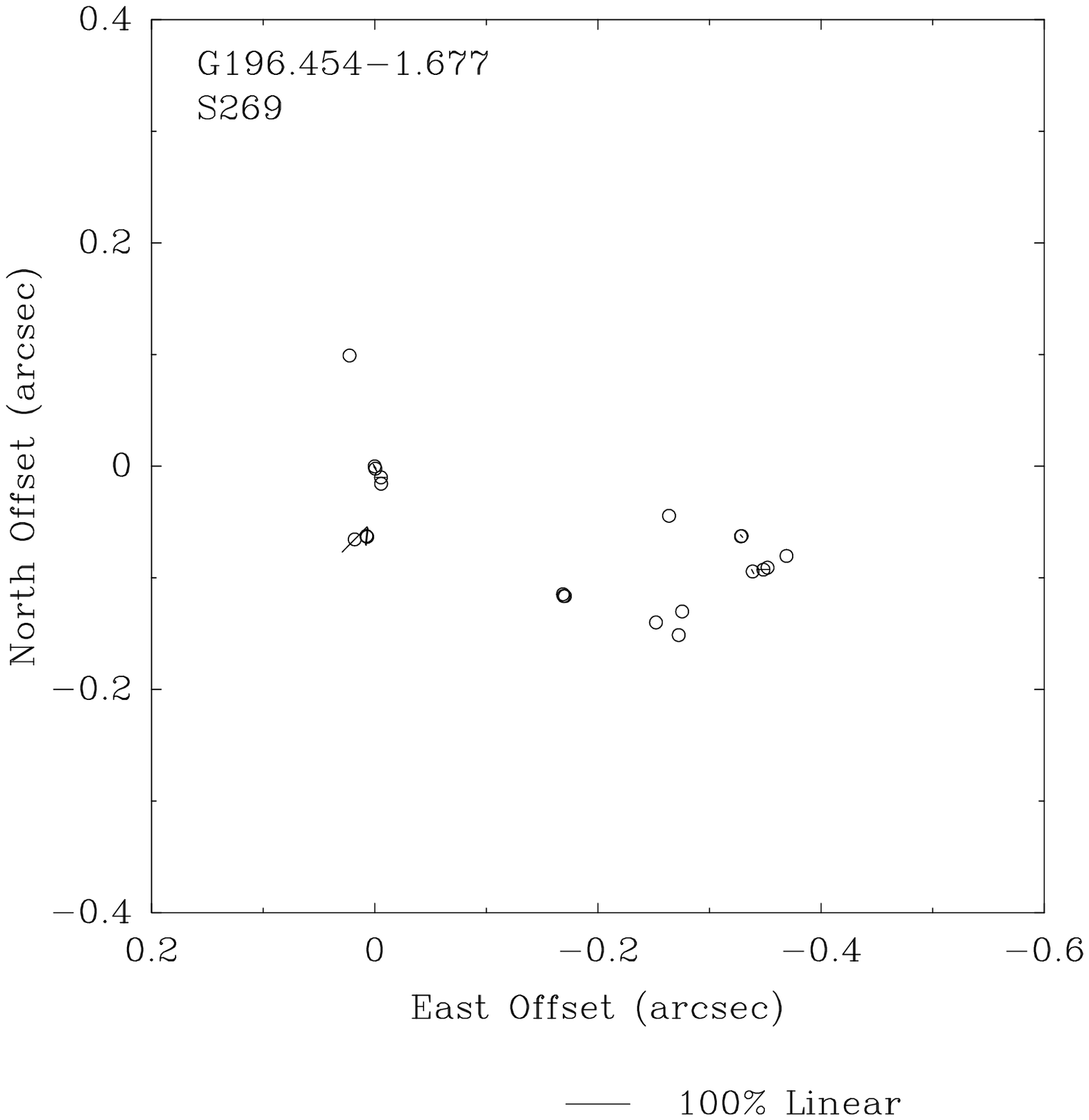}
\end{center}
\singlespace
\caption[Polarization map of S269.]{Polarization map of S269.  There
is no detectable continuum emission in the region.  Symbols are as in
Figure \ref{g5p}.\label{s269p}}
\doublespace
\end{figure}

\begin{figure}
\begin{center}
\includegraphics[width=6.0in]{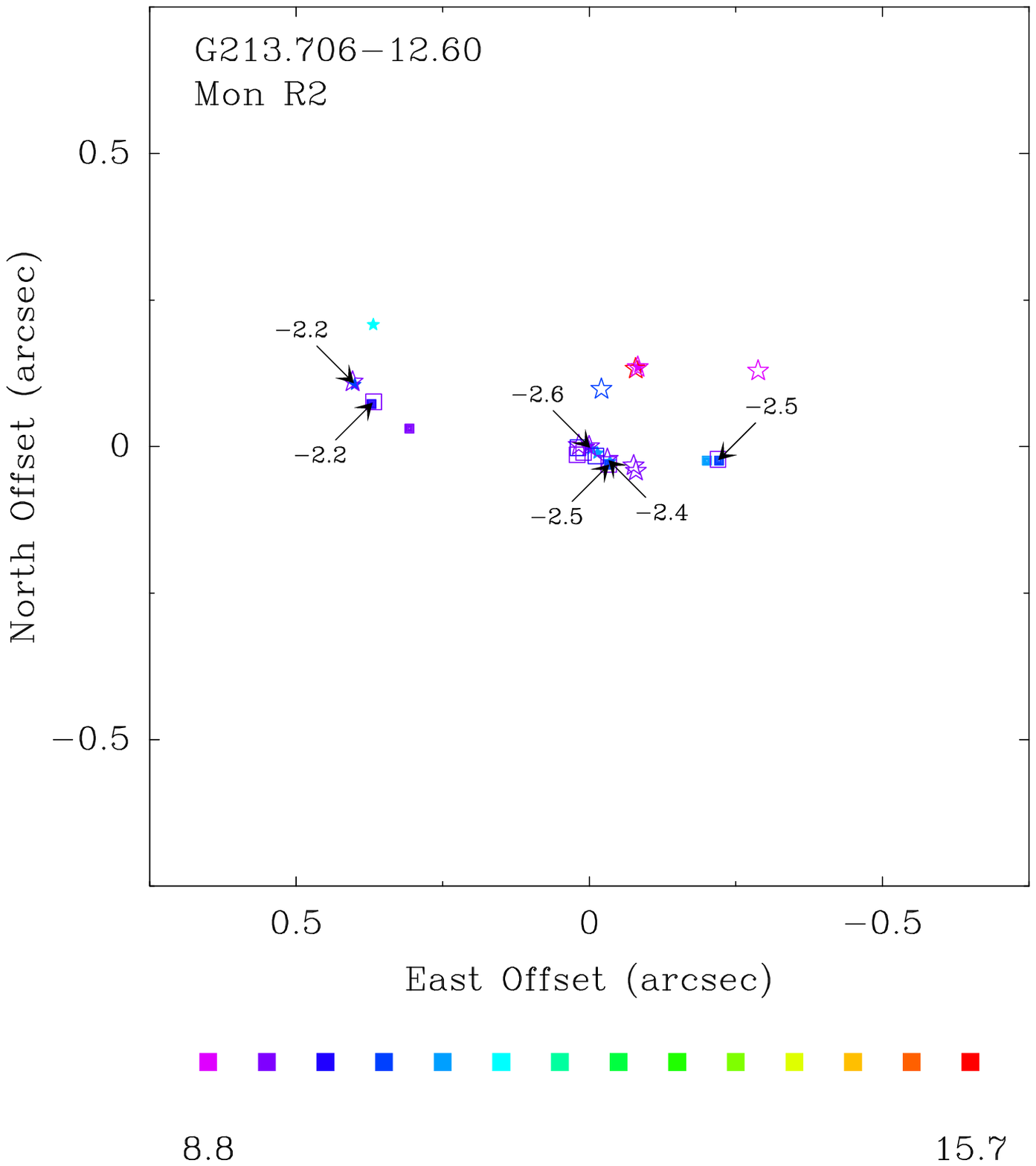}
\end{center}
\singlespace
\caption[Plot of maser spots in Mon R2.]{Plot of maser spots in Mon R2.
There is no detectable continuum emission in the region.  Symbols are
as in Figure \ref{g5v}.\label{monr2v}}
\doublespace
\end{figure}

\begin{figure}
\begin{center}
\includegraphics[width=6.0in]{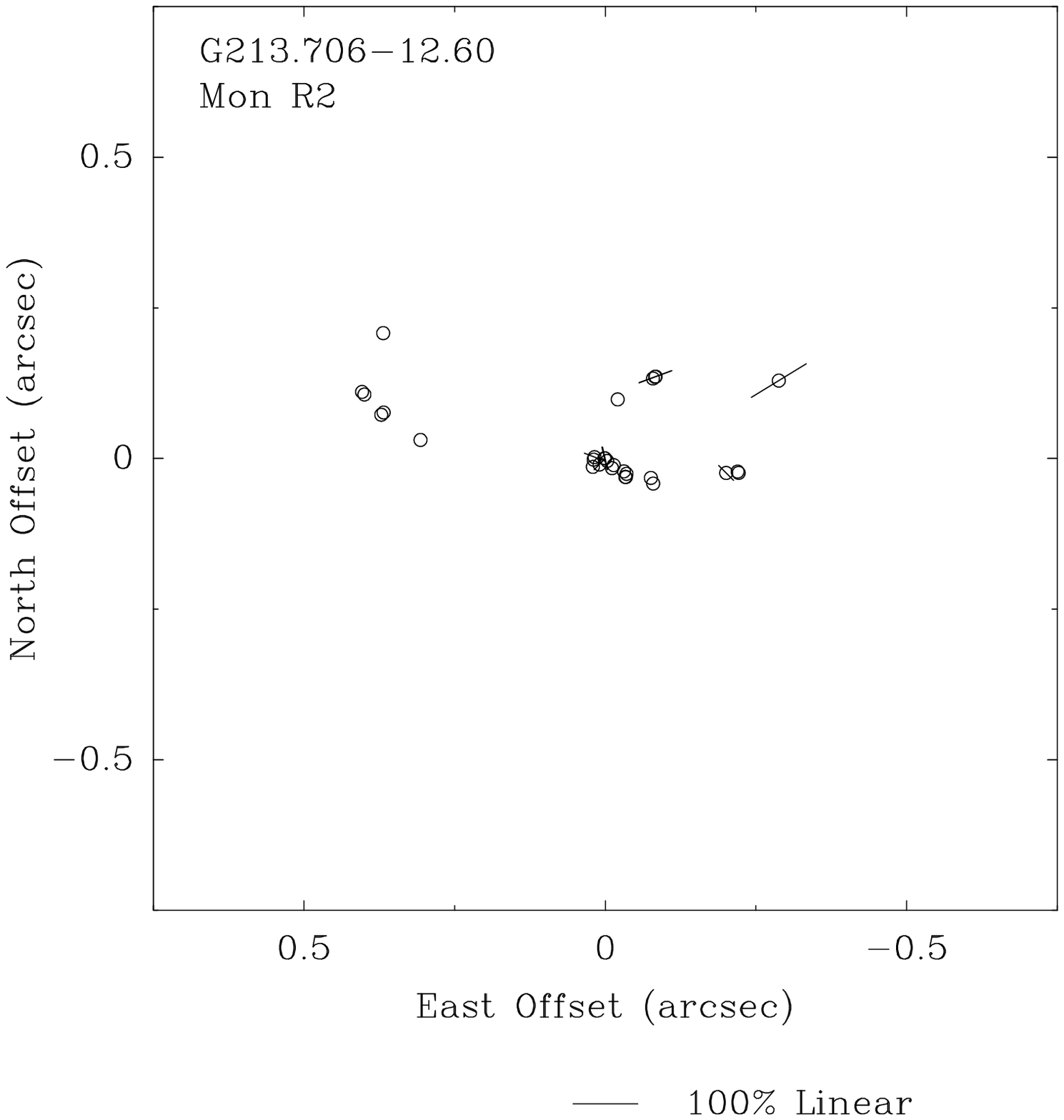}
\end{center}
\singlespace
\caption[Polarization map of Mon R2.]{Polarization map of Mon R2.
There is no detectable continuum emission in the region.  Symbols are
as in Figure \ref{g5p}.\label{monr2p}}
\doublespace
\end{figure}

\begin{figure}
\begin{center}
\includegraphics[width=6.0in]{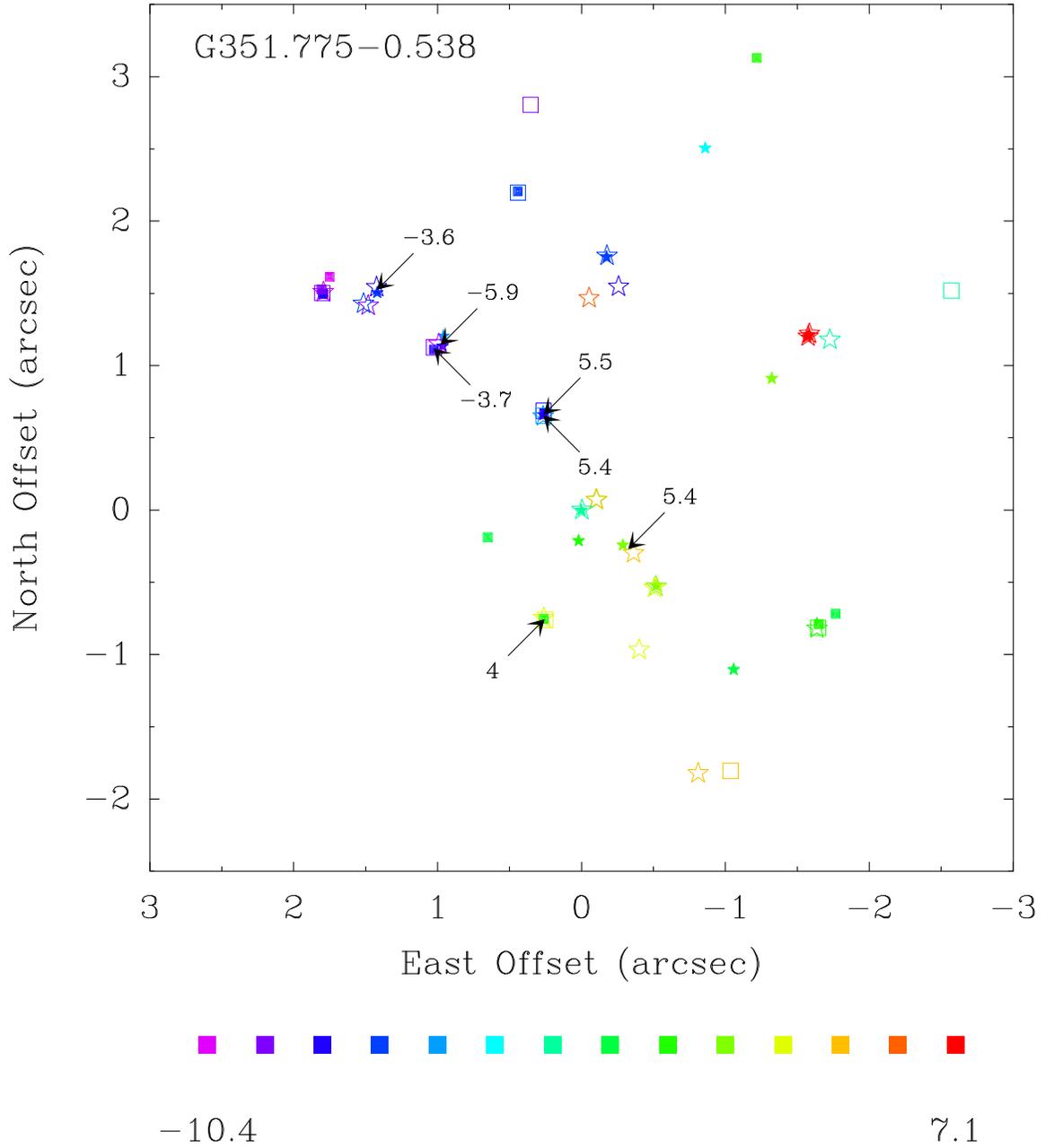}
\end{center}
\singlespace
\caption[Plot of maser spots in G351.775$-$0.538.]{Plot of maser spots
in G351.775$-$0.538.  Symbols are as in Figure \ref{g5v}.  A
continuum source to the east is not shown.\label{g351v}}
\doublespace
\end{figure}

\begin{figure}
\begin{center}
\includegraphics[width=6.0in]{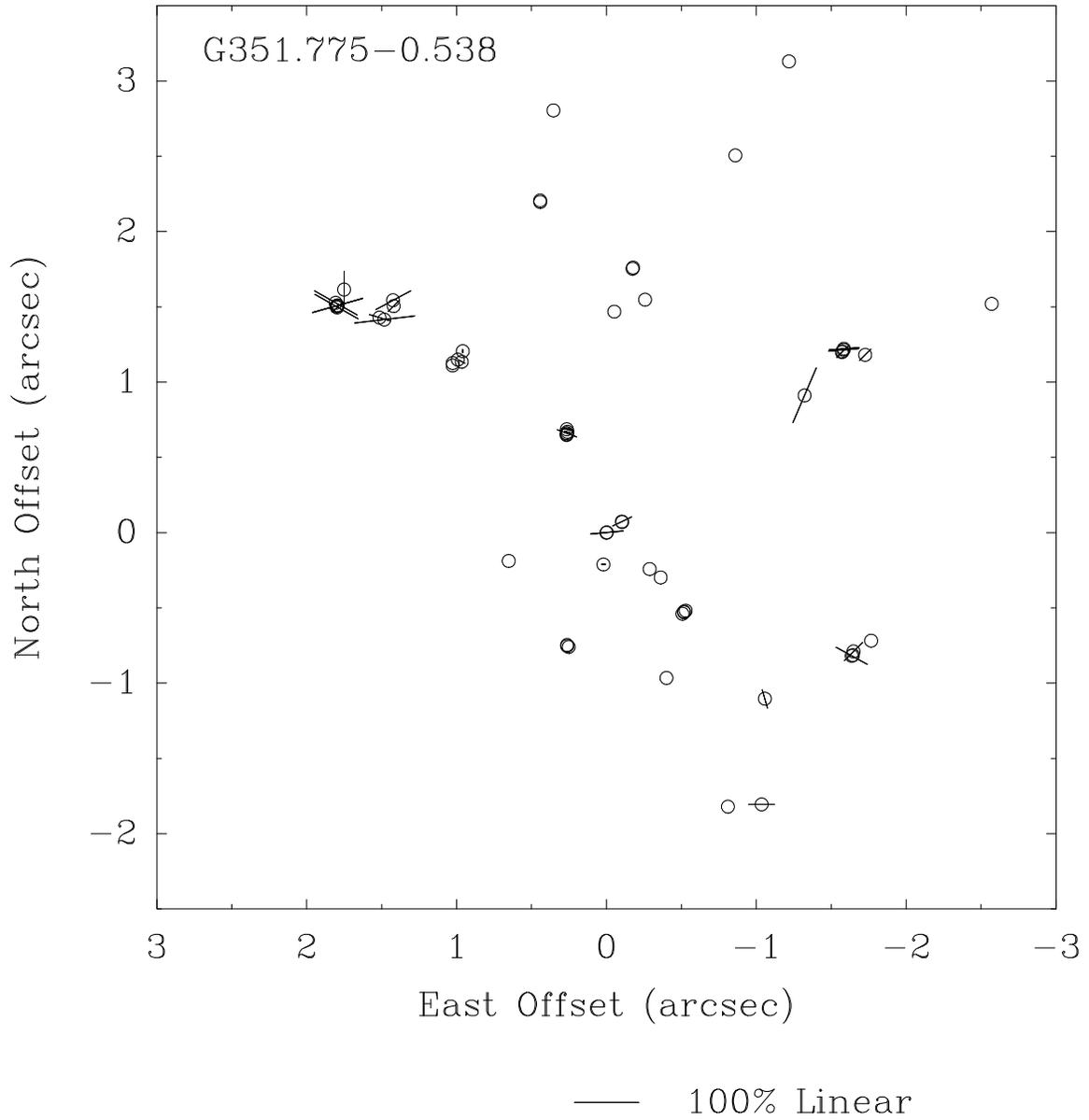}
\end{center}
\singlespace
\caption[Polarization map of G351.775$-$0.538.]{Polarization map of
G351.775$-$0.538.  Symbols are as in Figure \ref{g5p}.  A continuum
source to the east is not shown.\label{g351p}}
\doublespace
\end{figure}

The maser spots are shown superposed on continuum images in Figures
\ref{g5v} to \ref{g351p}.  Most of these are X-band continuum images
taken from the VLA survey \citep{arm}, although U-band continuum
images are substituted when they are of higher quality.  There is a
registration uncertainty between the maser spots and continuum maps.
We have attempted to use the same registrations as in the survey, but
this is not always possible for two reasons.  First, OH masers are
variable, so some maser spots have disappeared or appeared in the
years between the survey paper and the VLBA observations described in
this paper.  Second, the beamsize for the VLA OH observations was
greater than $1\arcsec$.  The VLA maps suffer from blending of maser
spots along with the concomitant positional smearing.  The positional
uncertainties of weak, isolated maser spots in the VLA survey are also
large.  Unlike in the VLA survey, however, there should be essentially
no relative registration error ($< 1$ mas) between the 1665 and
1667~MHz spots.  Both transitions were observed simultaneously, and
the self-calibration from one transition was applied to the other.
The maser spot used for self-calibration of each source is located at
the origin.

Maser spots are indicated by stars and squares for 1665 and 1667~MHz
emission, respectively in Figures \ref{g5v} to \ref{g351v}.  Emission
in RCP is represented by open symbols and LCP by filled symbols.
Locations with identified Zeeman pairs are indicated by arrows, and
the numbers indicate magnetic fields in milligauss.  Positive field
strengths denote fields for which the line-of-sight component is
directed away from the Sun ($v_{\mathrm RCP} > v_{\mathrm LCP}$),
while negative field strengths denote fields oriented toward the Sun.
The velocity of each maser spot is indicated in color, with a velocity
key appearing below each map.  The velocities are not corrected to
remove the Zeeman effect.  That is, a Zeeman pair consisting of two
$\sigma$-components at the same location would be indicated by an open
and filled spot at different velocities, even though the true velocity
of the material would be at the average of the two velocities.  While
the velocity of the material is more physically significant than the
uncorrected velocities of the maser spots, Zeeman velocity shifts have
not been removed because it is not possible to do so consistently for
all maser spots in a source.  Magnetic field strengths (and
directions) vary across sources, and in most sources the majority of
maser spots cannot be identified as components of a Zeeman pair (or
triplet).

Maser line parameters are also listed in Tables \ref{g5t} to
\ref{g351t}.  For each maser line, the position, velocity and velocity
FWHM, flux density, linear polarization parameters, spot fit
parameters, and minimum peak brightness temperature are given.  Flux
densities are given for RCP (Stokes RR), LCP (LL), and linear
($\sqrt{\mathrm{Q}^2 + \mathrm{U}^2}$) polarizations.  Positions and
fit parameters for maser lines seen in linear polarization are based
on a flux-weighted average of Stokes Q and U.  When the same
maser line is seen in more than one polarization, as is the case for
spots with a high degree of linear polarization or a substantial
unpolarized component, all relevant parameters except the linear
polarization fraction and position angle are given for each
polarization, and the entries are grouped together on adjacent lines
without additional interline spacing.  Thus, any maser line is
represented by one to three lines of the table, depending on its
polarization characteristics.  The listed linear polarization
parameters are based on brightnesses in the peak channel of brightest
circular polarization.  Since polarization properties may change
across a spot size, the listed linear polarization fraction may not
always be identical to that which would be computed from the listed
flux densities.

Since only two adjacent detectable channels of emission were required
for the identification of maser spots, it is not always possible to
fit for the velocity center and FWHM.  When such fits are impossible,
the velocity listed is the center velocity of the channel of peak
emission, and no FWHM is given.  The listed velocity may be offset
from the actual line velocity by up to half of the 0.176~\kms\
velocity channel width, although in most cases the velocity error
should be significantly less.

Generally the detectability limit for maser spots in a single channel
of the Stokes Q and U maps was 30-40~mJy/beam.  When no emission was
detected in either Stokes Q or U, the linear polarization fraction is
listed as zero.  This is provided in preference to an upper limit of
linear polarization in order to leave the tables less cluttered and
more easily readable.  Frequently a maser spot was seen in either
Stokes Q or U, but not both.  This corresponds to electric-vector
polarization position angles (PPAs) of 0, 45, 90, or 135$\degr$.  In
each case, the undetected Stokes parameter was assumed to be
identically zero, but could actually be nonzero within the previously
quoted noise limits.  In almost no cases would this lead to an error
of more than a few degrees in the quoted PPA.

\subsection{Polarization Maps}

Polarization maps of all the sources are shown in Figures \ref{g5p} to
\ref{g351p}.  Each maser spot is marked as a circle, and the electric
polarization vector is indicated with a line.  A line indicating 100\%
fractional linear polarization is shown below each map.  The ratio of
the length of the vector drawn through a maser spot and the length of
the vector shown in the key is the fractional linear polarization for
each spot.

\subsection{Zeeman Pairs}

Zeeman pairs were identified where an LCP and an RCP line are
coincident to within $10^{15}$~cm (67 AU).  These pairs are listed in
Table \ref{zeemantable}. Zeeman pairs with larger separations were
also identified when unambiguous.  For instance, the separation
between the components of the first listed pair for G351.775$-$0.538
is over 200~AU, but the lack of any other nearby maser spots leads us
to identify these spots as a Zeeman pair.  We do not attempt to
identify Zeeman pairs when the implied magnetic field would be less
than 0.5~mG.

\subsection{Ammonia Spectra}

\begin{figure}
\begin{center}
\includegraphics[width=6.0in]{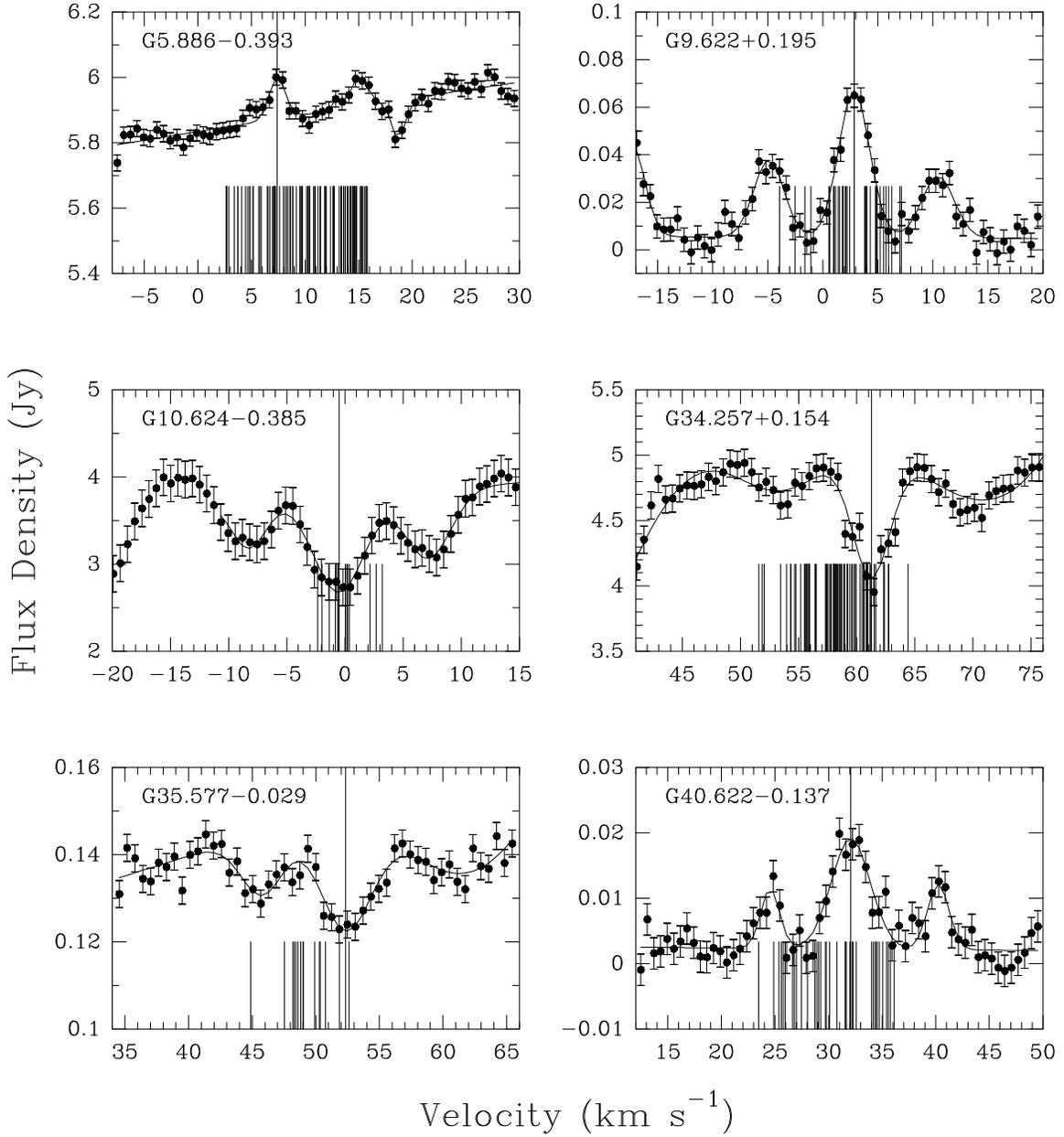}
\end{center}
\singlespace
\caption[Ammonia spectra.]{Spectra of (1,1) transition of NH$_3$ for
six sources.  Data points are shown with approximate errorbars.
Linear or quadratic baselines were fitted, and then Gaussians were
fitted the absorption or emission.  The velocity of the main hyperfine
transition is indicated with a vertical bar across each box.  OH maser
velocities are plotted as shorter vertical lines at the bottom of each
box.  Only the OH masers in Insets 1 and 2 of Figure \ref{g9v} are
shown in the upper right plot.\label{amm1}}
\doublespace
\end{figure}

\begin{figure}
\begin{center}
\includegraphics[width=6.0in]{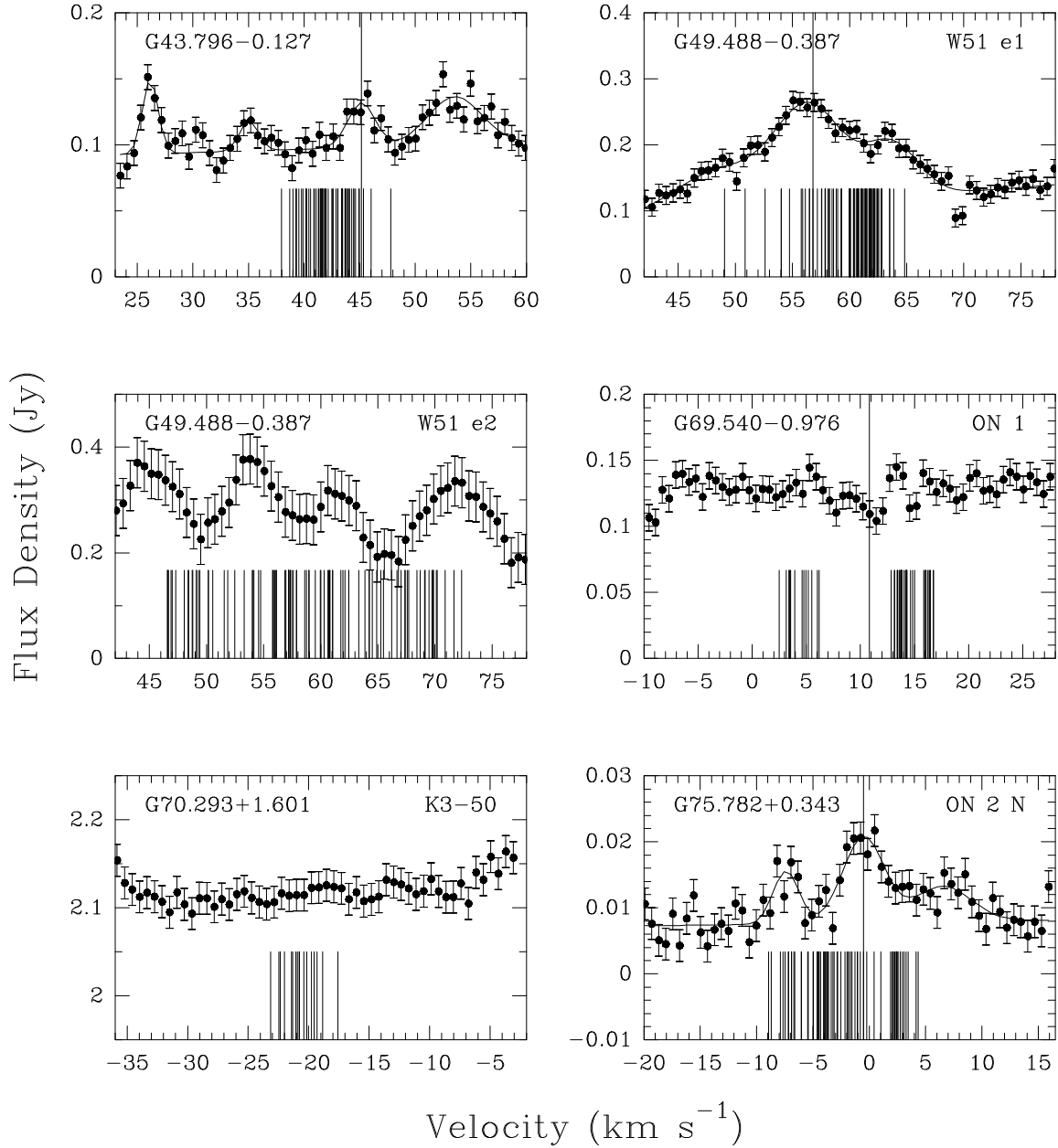}
\end{center}
\singlespace
\caption[Ammonia spectra.]{Spectra of (1,1) transition of NH$_3$ for
six sources.  See Figure \ref{amm1} for details.  Ammonia
emission at 10.8~\kms\ is seen in a large region 5\arcsec\ to
30\arcsec\ north of ON1.\label{amm2}}
\doublespace
\end{figure}

\begin{figure}
\begin{center}
\includegraphics[width=6.0in]{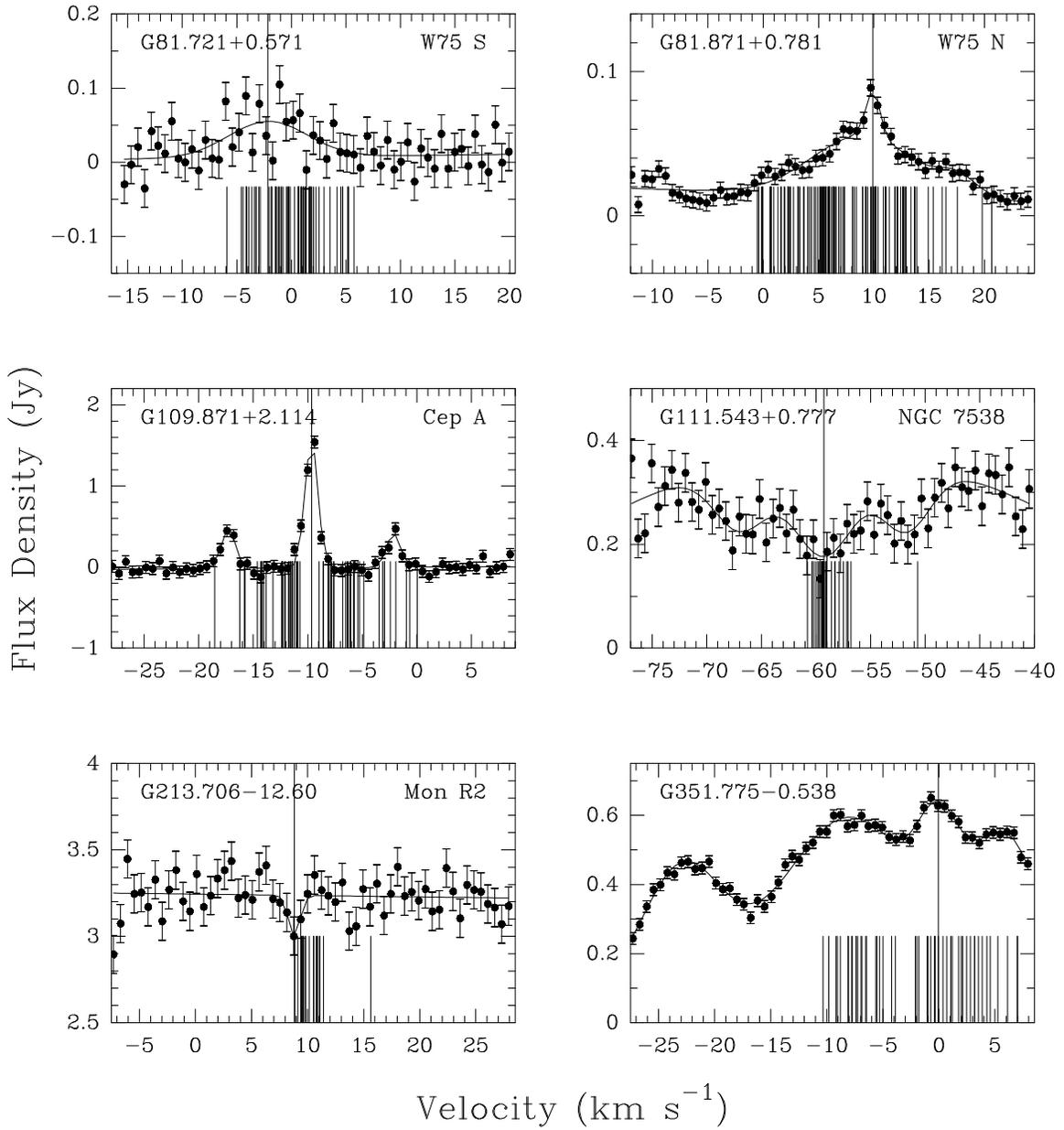}
\end{center}
\singlespace
\caption[Ammonia spectra.]{Spectra of (1,1) transition of NH$_3$ for
six sources.  See Figure \ref{amm1} for details.\label{amm3}}
\doublespace
\end{figure}

Spectra of ammonia emission or absorption are shown in Figures
\ref{amm1} to \ref{amm3}.  A vertical line running the entire height
of the box represents the velocity of the main hyperfine transition.
No such line is drawn in cases where no clear absorption or emission
is seen or in cases where it is unclear which line is the main
hyperfine line.  The velocities of the OH masers are shown as shorter
vertical lines at the bottom of each panel.  Analysis of the
implications of the relation of OH maser velocities to ammonia
velocities will be provided in a companion paper.

\singlespace

\clearpage

\begin{figure}
\begin{center}
\includegraphics[width=6.0in]{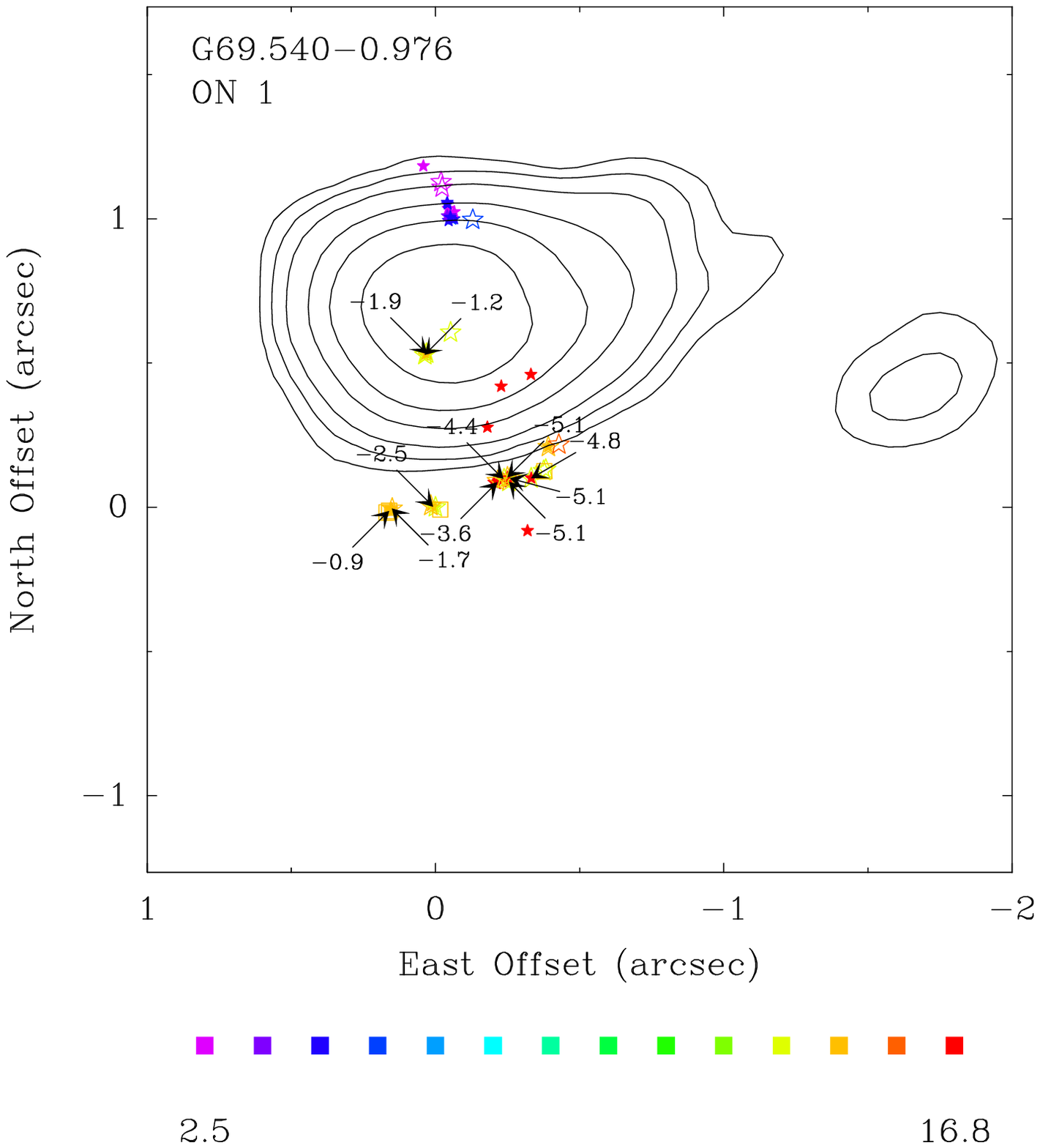}
\end{center}
\singlespace
\caption[Plot of maser spots in ON 1.]{Plot of maser spots in ON 1.
Symbols are as in Figure \ref{g5v}.}
\doublespace
\end{figure}

\clearpage

\clearpage

\clearpage

\notetoeditor{The spacing is intentional in the following tables.
  When the same maser feature is seen in more than one polarization
  (RCP, LCP, or Linear), it is given an entry in the table for each
  polarization in which it is detected, and all these lines are
  grouped together with smaller interline spacing.  Thus, I have
  defined two macros: backslash-sk to skip (add extra vertical space)
  between maser features, and backslash-ns to not skip (add negative
  vertical space) between table lines associated with the same maser
  feature.  If it is necessary to alter the table layout for
  publication, please do so in a way that maintains this distinction.}



\renewcommand{\arraystretch}{1.0}

\clearpage
\doublespace

\appendix
\section{Full-Polarization Spectral-Line VLBI Reduction Techniques}

The following procedure was used to reduce our full-polarization,
spectral-line, VLBA data.  Names of AIPS tasks and runfiles
corresponding to each step are given in parentheses.  Note that
simplified interfaces to some of these tasks are available as a series
of AIPS runfiles.\footnote{AIPS Memo 105 \citep{ulvestad} is available
at: ftp://ftp.aoc.nrao.edu/pub/software/aips/TEXT/PUBL/AIPSMEM105.PS .}

The full data set is loaded (\texttt{FITLD}), discarding data with low
weights.  Here and throughout the procedure, any obviously bad data
are flagged (\texttt{UVFLG}).  It is necessary to correct for digital
sampler bias by the VLBA correlator (\texttt{ACCOR}).  \emph{A priori}
amplitude calibration is applied (\texttt{APCAL}) using the gain curve
and system temperature tables provided with the data set.  Since VLBA
antennas have altitude-azimuth mounts, the parallactic angle of a
polarized source will change with time due to the relative geometry
between the source and antenna.  It is necessary to remove these
effects (\texttt{CLCOR, OPCODE='PANG'}) for full-polarization
observations.

Instrumental phases and single-band delays are removed by fringe
fitting (\texttt{FRING}) on a good scan of a strong calibrator.  Fringe
fitting is computationally intensive and requires significant amounts
of computer memory, so it may be advisable to average several spectral
channels together (\texttt{AVSPC}).  Fringe rates should be zeroed
(\texttt{SNCOR, OPCODE='ZRAT'}) before being applied to the spectral
line source, since the rates determined from the calibrator may worsen
phase coherence.  For full-polarization observations, it is also
necessary to determine the delay difference between RCP and LCP data
(\texttt{CRSFRING}).

The amplitude bandpass response should be obtained from a strong
calibrator source using autocorrelation data (\texttt{BPASS}).  The
bandpass should be examined (\texttt{BPLOT}) to ensure that bad data
and interference have not corrupted the output.  If multiple
calibrator scans are used to solve for the bandpass response, the
bandpass can be interpolated in time to correct for small,
time-varying instrumental effects.

For high-precision polarimetry it is necessary to determine the
instrumental calibration.  The feed D-terms add vectorially to the
source polarization in the Stokes $(Q, iU)$-plane.  Since the
parallactic angle calibration has already been applied, the source
polarization is constant, but the instrumental polarization varies
with time.  Thus, it is possible to solve for this polarization if an
unpolarized source has been observed over a wide range of parallactic
angles.  The data from a strong, unpolarized calibrator should be
self-calibrated (\texttt{CALIB, IMAGR}) before solving for the D-terms
(\texttt{LPCAL}).  Instrumental polarization terms at the VLBA are
typically on the order of a few percent.

Next, the phase difference between RCP and LCP must be calculated,
since this corresponds to a rotation of the polarization position
angle.  A strong calibrator with known polarization characteristics
should have been observed.  The data for this source should be
self-calibrated and imaged in Stokes Q and U, and the resulting
polarization vectors should be compared against the known
polarization.  For 3C286 in AIPS, the applicable polarization angle
correction is $66\degr - \mathrm{arctan}(U,Q)$, where a two-argument
arctangent function is used.

Spectral lines drift back and forth in frequency as their
Doppler-shifted velocity changes with time over the course of
observations due to the rotation of the Earth around the Sun.  In
AIPS, these effects can be removed by setting the LSR velocity and
rest frequencies of the observations (\texttt{SETJY}) and then running
\texttt{CVEL}.

The maser sources can be self-calibrated using a bright maser spot as
the reference feature.  The resulting calibration can be applied to
all spectral channels in both polarizations.  Experience has shown
that this calibration can be applied to the other main-line,
ground-state OH transition as well if both are observed
simultaneously.  While the absolute position of maser spots cannot be
obtained using self-calibrated data, the relative positions of all
spots in both polarizations in both transitions will be correct.
After self-calibration, the previously obtained RCP $-$ LCP phase
difference should be applied (\texttt{CLCOR, OPCODE='POLR'}) before
final imaging.

\end{document}